\documentclass[aip,reprint,pre]{revtex4-2}
\usepackage{epsfig,graphicx}
\usepackage{amssymb,amsfonts,amsmath,bm}
\usepackage{hyperref}

\begin{document}
	\title{Nonequilibrium Optimal Reaction Coordinates for Diffusion.}
	
	\author{Sergei V. Krivov}
	\email{s.krivov@leeds.ac.uk}
	
	\affiliation{University of Leeds, Astbury Center for Structural Molecular Biology, Faculty of Biological Sciences, University of Leeds, Leeds LS2 9JT, United Kingdom}

\begin{abstract}
Complex multidimensional stochastic dynamics can be approximately described as diffusion along reaction coordinates (RCs). If the RCs are optimally selected, the diffusive model allows one to compute important properties of the dynamics exactly. The committor is a primary example of an optimal RC. Recently, additive eigenvectors (addevs) have been introduced in order to extend the formalism to non-equilibrium dynamics. An addev describes a sub-ensemble of trajectories of a stochastic process together with an optimal RC. The sub-ensemble is conditioned to have a single RC optimal for both the forward and time-reversed non-equilibrium dynamics of the sub-ensemble. Here we consider stationary addevs and obtain the following results.  We show that the forward and time-reversed committors are functions of the addev, meaning a diffusive model along an addev RC can be used to compute important properties of non-equilibrium dynamics exactly. The rates of the conditioned stochastic process, describing an addev sub-ensemble, can be computed ``on the fly'', thus allowing efficient sampling schemes. We obtain two families of addev solutions for diffusion. The first is described by equations of classical mechanics, while the second can be approximated by quantum mechanical equations. We show how the potential can be introduced into the formalism self-consistently. The developments are illustrated on simple examples. The two addev families provide stochastic models together with optimal RCs and new sampling schemes for classical and quantum mechanical systems. It suggests that the conditioned stochastic processes defined by addevs are suitable for description of dynamics of physical origin.



\end{abstract}
\maketitle

\section{Introduction}
A fundamental approach to analyse complex multidimensional stochastic dynamics, in particular that of a chemical reaction, is to project it onto one or a few reaction coordinates (RCs) \cite{du_transition_1998, best_reaction_2005, coifman_diffusion_2006,nadler_diffusion_2006,altis_dihedral_2007,krivov_diffusive_2008,peters_reaction_2013,peters_reaction_2016,banushkina_optimal_2016}. Free energy and diffusion coefficient as functions of the employed RCs define a diffusive model of the projected dynamics. The model describes the dynamics projected on the RCs as diffusion on the free energy landscape, which provides a simple, visually appealing picture of the overall dynamics \cite{banushkina_optimal_2016}. In particular, it can be used to locate the top of the free energy barrier or the transition state as the rate-limiting step or the bottleneck of the chemical reaction. For such a model to provide an accurate, quantitative description of the dynamics, the RCs should be chosen in an optimal way, e.g., in order to minimize non-Markovian effects due to the projection \cite{banushkina_optimal_2016,krivov_is_2010}. For equilibrium reaction dynamics between two end states $A$ and $B$ such an optimal RC is know as the committor or splitting probability and equals $q_B(x)$ - the probability to reach state $B$ before reaching state $A$ starting from configuration $x$ \cite{onsager_initial_1938,du_transition_1998}. The diffusive model along the committor can be used to compute \textit{exactly} the following important properties of the dynamics - the equilibrium flux, the mean first passage times (mfpt) and the mean transition path times (mtpt) between any two points on the committor \cite{krivov_reaction_2013,berezhkovskii_diffusion_2013,lu_exact_2014,krivov_protein_2018}. This is true for free energy landscapes of any complexity and does not require separation of time-scales. The diffusive model can be used to determine accurately and in a direct manner the free energy barrier and the pre-exponential factor - the two major determinants of the reaction dynamics \cite{krivov_protein_2018}. Note that it is generally much more difficult to compute these properties using alternative approaches, e.g,  the Markov state models \cite{chodera_markov_2014,schwantes_improvements_2013,krivov_blind_2020}.

The committor is a complex, high-dimensional function, which accurate determination for realistic systems of interest, e.g., protein folding trajectories, is a difficult task \cite{krivov_hidden_2004,freddolino_challenges_2010}. A large number of methods have been developed to accurately determine the committor or other optimal RC \cite{best_reaction_2005,  peters_obtaining_2006, ma_automatic_2005, lechner_nonlinear_2010, perez_identification_2013,  elber_calculating_2017, jung_artificial_2019, hernandez_variational_2018, khoo_solving_2018, li_computing_2019, mardt_vampnets_2020, he_committor_2022, hasyim_supervised_2022, banushkina_nonparametric_2015, krivov_protein_2018, krivov_blind_2020, krivov_nonparametric_2021, li_computing_2019, frassek_extended_2021, lucente_coupling_2022, chen_committor_2023}. We have suggested a non-parametric framework for the determination and validation of optimal RCs \cite{banushkina_nonparametric_2015, krivov_protein_2018, krivov_blind_2020, krivov_nonparametric_2021}. In contrast to alternative (parametric) approaches, which require a functional form with many parameters to approximate an RC and thus extensive expertise with the
system, the developed approaches are non-parametric and can approximate any RC with high accuracy without system specific information. The developed validation/optimality criteria can be used to validate that the putative RC is indeed optimal or if not, then identify the most sub-optimal regions of the RC. The developed approaches were successfully tested on realistic protein folding trajectories. The optimal RCs, determined with the non-parametric approaches, have passed the stringent validation criteria. Recently, the framework has been extended to non-equilibrium/adaptive sampling \cite{krivov_nonparametric_2021}, e.g., a large ensemble of short trajectories - a promising approach toward simulations employing exascale or cloud computing \cite{kohlhoff_cloud_2014,lohr_abeta_2021}.

Originally, the free energy landscape and optimal RC frameworks were introduced/developed for the description of chemical reaction dynamics. However, such a description, in principle,  can be applied to stochastic dynamics other than chemical reactions, where the two boundary states can be defined. For example, it can be applied to describe the dynamics of a disease, with two natural boundary states being completely healthy and very ill or "dead"; it has been applied to describe dynamics of patient recovery after kidney transplant \cite{krivov_optimal_2014}. The approach has been applied also to the game of chess \cite{krivov_optimal_2011}.

While, no doubt, very useful, the description of stochastic dynamics with the committors as RCs has major shortcomings. 1) The committor function requires the specification of two boundary states. The proper specification of boundary states for complex systems, e.g., protein folding trajectories is a difficult task \cite{krivov_blind_2020}. For systems with deep free energy basins and a high free energy barrier one can use the slowest eigenvector to define the two regions of configuration space that belong to the two minima. However, this is likely to fail for system with many shallow free energy minima, e.g.,  natively unstructured proteins. For some system, e.g., a harmonic well, introduction of boundary states makes no sense at all. A related shortcoming is that the committor RC describes the dynamics just between the boundary states, i.e., dynamics of transition paths (TP). The dynamics inside the boundary states is not described. While dynamics of TPs is very important, and arguably, is sufficient for the understanding of reaction, one may want to describe the entire dynamics/trajectory. Such a description is even more important for systems without well-defined boundary states, e.g., natively unstructured proteins or a harmonic well. 2) For the description of non-equilibrium (without the detailed balance) reaction dynamics between two boundary states one needs to consider two committors $q_B$ and $q'_A$, where the latter is the committor for the time-reversed dynamics \cite{metzner_transition_2009}. The two committors are different functions and how one obtains a single optimal RC out of them is not clear. In principle, the knowledge of the two committors is sufficient to compute the equilibrium flux, the mfpt and the mtpt between the boundary states. However, unlike the equilibrium case, the two committors can not be used to compute these properties for intermediate boundaries because their iso-committor surfaces do not coincide. 

It is important to extend/generalize the framework of optimal RCs further, for the analysis/description of generic stochastic dynamics, e.g., stochastic dynamics without boundary states, or non-equilibrium stochastic dynamics. A practically important case of the latter are trajectories of molecular dynamics simulations observed with relatively short lag times, when the system still remembers its momenta. Analysis of such trajectories at short lag times should allow significantly shorter trajectories and higher efficiency in parallel approaches for exascale computing, which use a very large ensemble of short trajectories instead of a single long one \cite{lohr_abeta_2021,kohlhoff_cloud_2014,zhou_fegs_2012,doerr_learning_2014,wan_adaptive_2020,perez_adpative_2020,pan_string_2008,lev_string_2017}.

Recently we have suggested the additive eigenvectors (addev) as optimal RCs \cite{krivov_method_2013,krivov_addev_2022}, which are free from the shortcomings associated with the committors. In particular, a single optimal RC (the phase of an addev) describes both the forward and time-reversed non-equilibrium dynamics. They are found as eigenvectors of the addev master equation (AME) and dont require the specification of boundary states. The AME provides a time-reversible description of stochastic dynamics, i.e., the probability evolution can be computed forward and backward in time. Here we focus on stationary solutions of the AME for diffusion. The solution, an addev, describes stochastic dynamics as a stochastic eigenmode - stochastic periodic processes. For example, while the committor describes the TP segments of a protein folding reaction, an addev should describe folding reaction as a periodic process where a protein repeatedly folds and unfolds. 

\begin{figure}[htbp]
	\centering
	\includegraphics[width=.8\linewidth]{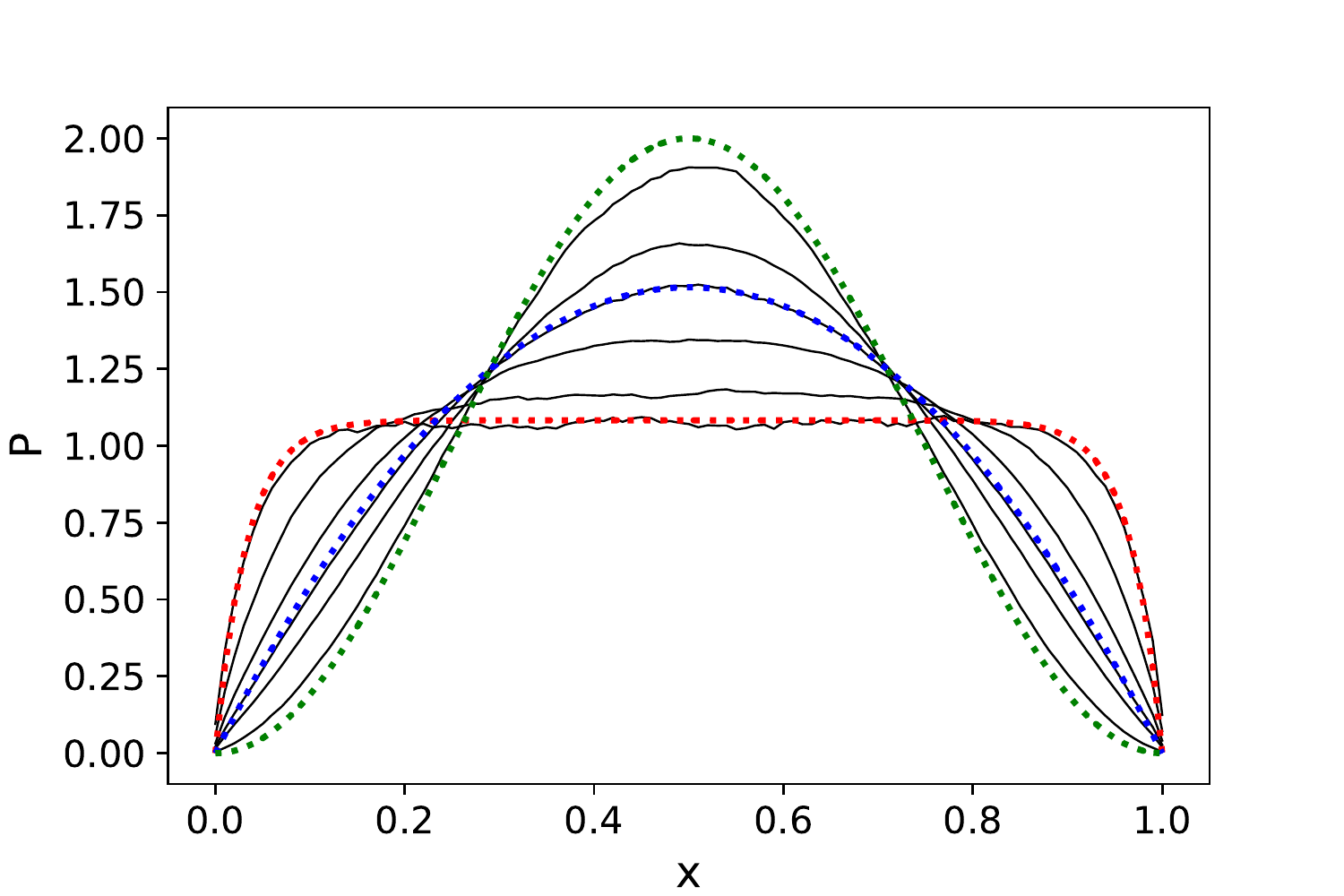}
	\caption{TPs for free diffusion on the interval $0\le x\le 1$. Thin black lines show stationary distribution of TPs conditioned on duration $\tau$ for (from bottom to top) $0< \tau<0.05$,  $0.05< \tau<0.1$, $0.1<\tau<0.2$, $0.2<\tau<0.25$, $0.25<\tau<0.5$, $0.5<\tau<1$, $1.4<\tau$. Dotted lines show that for TPs computed for addev sub-ensembles: almost constant (red), parabolic $6x(1-x)$ (blue), and quasi-stationary distribution $2\sin^2(\pi x)$ (green); for details see text.}
	\label{fig:TPtau}
\end{figure}

Addevs describe stochastic dynamics differently compared to standard approaches, e.g., the master equation or the Fokker-Planck equation \cite{risken_fokker-planck_1984}. They describe dynamics of sub-ensembles of trajectories, conditioned on having the same optimal RC for the forward and time-reversed dynamics. It means that one may need to consider all addevs, which contribute to the process of interest. To illustrate this on a specific example and to show how description with addevs is related to standard description of the TPs with the committors consider a simple example of drift free one-dimensional diffusion, with the diffusion coefficient $D(x)=1$, on the interval $0\le x \le 1$. Introduce two boundary states $x=0$ and $x=1$ and consider the TP sub-ensemble of trajectories, i.e., segments of trajectories that start in one boundary state and end in the other without visiting a boundary state in-between. We reserve the word ensemble for the entire ensemble of trajectories. Sub-ensemble refers to a sub-set of trajectories selected according to some rule. The properties of the TP sub-ensemble are described by the committor function $q_B(x)=x$. For example, the stationary distribution of the sub-ensemble is parabolic  $P(x|TP)=6x(1-x)$. Now, divide the TP sub-ensemble onto different sub-ensembles of TPs conditioned on having specific duration $\tau$. These sub-ensembles have different statistical properties, in particular different stationary distributions, as shown on Fig. \ref{fig:TPtau}. For duration $\tau$ around the typical, average value $\langle \tau \rangle$ the statistical properties are not perturbed and are similar to that of the entire TP sub-ensemble. However, for duration much shorter or much longer, they are different. Nevertheless, each such sub-ensemble can be approximated by an addev.  For very short durations $\tau$, the distribution is almost constant; corresponding to trajectories rapidly moving forward with constant drift. At the other extreme of large $\tau$, the stationary distribution approaches $2\sin^2(\pi x)$ - the quasi-stationary distribution of trajectories avoiding the end states $x=0$ and $x=1$. Indeed, to have a very long duration the trajectory should avoid the boundary states. These different sub-ensembles of trajectories correspond to different addevs. As we show below, one can compute forward and time-reversed committors and describe TPs in the addev sub-ensemble, in particular their stationary distribution. The red dotted line shows the stationary TP distribution in an addev with strong forward bias. The blue dotted line corresponds to an addev with no bias, which reproduces the standard parabolic $6x(1-x)$ distribution. The quasi-stationary distribution, shown by the green dotted line, corresponds to another addev solution for a bound state on an interval. Thus instead of a single TP sub-ensemble we have a collection of addev sub-ensembles, each with its own conditioned dynamics and corresponding optimal RC. In this simple one-dimensional example all the optimal RCs are the same, however in a general case of non-equilibrium multidimensional dynamics different addevs generally have different optimal RCs and describe different aspects of dynamics. 

The advantage of such an expanded (or over-complete) description can be seen as follows. Consider the description of reaction dynamics with the committor. The diffusive model along the committor can be used to compute exactly, for example, the mtpt; however higher moments or the entire distribution can not be computed very accurately unless there is a separation of time scales and dynamics, projected on the committor, is Markovian. Assume now that the entire ensemble of all trajectories is divided into sub-ensembles of similar trajectories, for example, according to different pathways. Then one can find committor for each such sub-ensemble and the corresponding diffusive model would provide much more accurate description of each such sub-ensemble. The entire ensemble of trajectories can be approximated by a combination or sum of such sub-ensembles with appropriate weights. Addevs provide analogous decomposition. Each addev describes a sub-ensemble of trajectories together with the corresponding optimal RC. The sub-ensembles, however, are organized not according to different pathways, as that would be system specific and would require some apriory information about the system, but according to different ways the dynamics can be biased under condition that the optimal RC for forward and time-reversed dynamics is the same.

We define addevs as sub-ensembles of trajectories of a conventional Markov chain, conditioned to have the same optimal RC for forward and time-reversed dynamics. These are biased or atypical trajectories of the Markov chain and the probability to observe a trajectory of the Markov chain in such an addev sub-ensemble exponentially decreases with time, i.e., it is a rare event. Thus, in order to approximate the dynamics of the Markov chain, one needs to consider a collection of addevs, as illustrated above. However, stochastic dynamics described by a single addev is interesting by itself, and can be used for tasks other then approximation of a standard Markov chain dynamics. Moreover, as we show below, due to gauge-invariance, addevs introduce potential in a different way compared to the standard Markov chains. It means that for stochastic dynamics with potential, they describe essentially different stochastic processes. Here we describe two families of stationary addev solutions for diffusion and show that the first family is described by equations of classical mechanics, while the second family can be accurately approximated, in some regimes, by quantum mechanical equations. We show that rates of the Markov chain describing conditioned dynamics in an addev sub-ensemble can be computed ``on the fly'', allowing efficient sampling of addevs, in particular, for stochastic dynamics with potential. Summarizing, addevs provide stochastic models together with optimal RCs and efficient sampling schemes of classical and quantum mechanical systems; for the latter, in the regimes where the approximation is close.

The paper is as follows. Section \ref{sec:theory} presents new fundamental developments. Here we start with introducing the addevs and AME, and deriving some addevs properties. Namely, that the forward and time-reversed committors can be expressed as functions of addevs, establish the gauge invariance, present expressions to compute various properties of the addev sub-ensembles and discuss how addev equations can be solved ``on the fly''. Section \ref{sect:classical} describes the first family of addev stationary solutions for diffusion. We derive a PDE describing such solutions, which is equivalent to the classical Hamilton-Jacobi equation and provide an SDE for sampling of such an addev dynamics. A few illustrative examples are presented. Section \ref{sect:potential} describes how to properly introduce potential into the addev formalism and the AME. Section \ref{sect:idf} presents the second  family of addev stationary solutions for diffusion, addevs with an internal degree of freedom, which provide a different description of the same diffusion dynamics. We show that in some regimes the solutions can be accurately approximated by that of the quantum mechanical equations. Illustrative examples are presented at various level of details. We conclude with a discussion.

\section{Theory}
\label{sec:theory}
In this section we briefly introduce the additive eigenvectors (addevs), the corresponding diffusive model along the optimal RC, and the addev master equation (AME), derived in Ref.  ~\citenum{krivov_addev_2022}.
\subsection{Addevs and nonequilibrium diffusive model}
\begin{figure}[htbp]
	\centering
	\includegraphics[width=.9\linewidth]{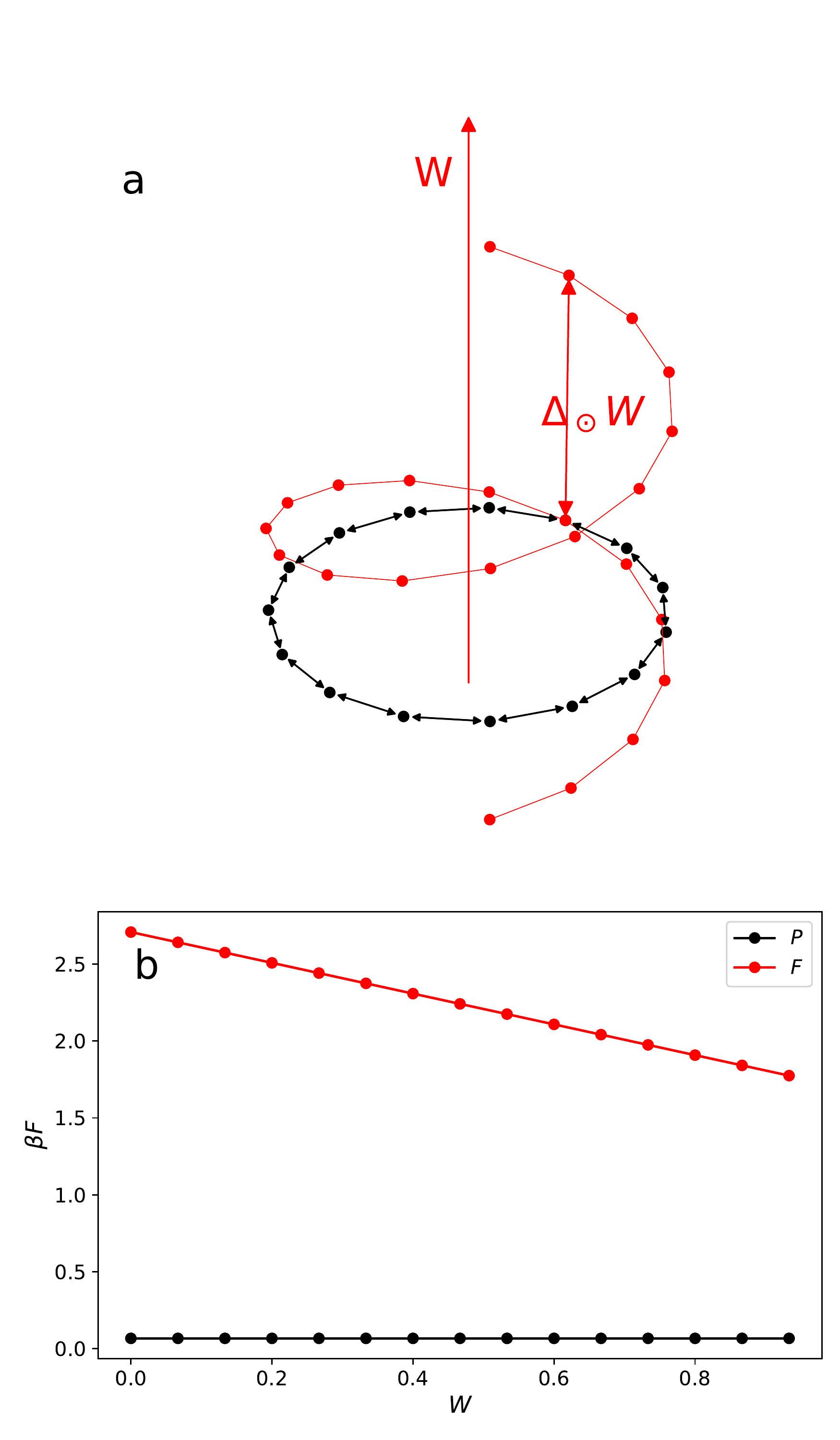}
	\caption{A typical diffusive model along $W$ for a simple Markov chain on a circle. \textbf{a)} Configuration space of the Markov chain ($N=15$ states) and the multivalued optimal RC $W$ shown in black and red, respectively. $\Delta_\odot W$ is the increment of $W$ after a complete revolution. \textbf{b)} Stationary probability (black) $P(W)=1/N$ and the free energy (red) of the diffusive model $F(W)$ as functions of RC $W$. $P(W)$ is single valued, while $F(W)$ is multivalued, according to Eq. \ref{Fw}. $F(W)$ is shown for one period of $W$, here $\Delta_\odot W=1$. $F(W)$ describes diffusion with a constant drift and a non-zero flux to the right on panel \textbf{b}, or rotation in the direction of increasing $W$ on panel \textbf{a}, i.e.,  a stochastic eigenmode.}
	\label{fig:FW}
\end{figure}

A diffusive model along an addev RC describes nonequilibrium diffusion dynamics with non-zero flux $J$. Since flux is constant and should be zero at a boundary, it means that an addev RC, unlike the committor, has no boundaries. Thus, an addev RC is either the infinite line or it is a multivalued function on a circle, similar to the angle (Fig. \ref{fig:FW}a). Equation for an addev optimal RC $W(i)$ for a Markov chain is obtained by considering frame of reference moving with the flux, $S(i,t)=W(i)-\nu t$ and requiring, analogous to the committor \cite{krivov_method_2013-1,krivov_addev_2022}, zero average displacement, i.e., $S$ is a space-time harmonic function   
\begin{equation}
	\label{S0}
	\sum_j P_\tau(j|i)[S(j,t+\tau)-S(i,t)]=0,
\end{equation}
here, $P_\tau(j|i)$ denotes transition probability from state $i$ to state $j$ after time interval $\tau$.
Eq. \ref{S0}, can be written symbolically for $W$ as $WP_\tau=W+\nu\tau$, meaning that action of matrix $P_\tau$ on vector $W$ does not change the vector and just adds a constant, which explains the term additive eigenvector with $\nu$ being the additive eigenvalue. Generally, $S$ (or $W$) are different for the forward and time-reversed dynamics. For such systems, as described in the next section, one considers sub-ensembles of trajectories conditioned on having the same $S$ (or $W$) for the forward and time-reversed dynamics. For dynamics in a such  sub-ensemble, the framework of optimal RCs can be applied \cite{krivov_reaction_2013,krivov_addev_2022}. In particular, the dynamics can be accurately approximated by a diffusive model, namely the free energy $F(W)$, and position dependent diffusion coefficient $D(W)$. The free energy $F(W)$ can be computed from the stationary probability distribution $P(W)$ as 
\begin{subequations}
	\label{Fw}
	\begin{align}
		\beta F(W)&=-\ln P(W)-\alpha W \\
		\alpha&=\nu/(\Delta_\odot W Z_{C,1}/T),
	\end{align}
\end{subequations}
where, $\beta$ is an inverse temperature, $Z_{C,1}(W)\sim P(W)D(W)$ is a cut-profile \cite{krivov_reaction_2013,krivov_addev_2022} which is constant along $W$, $T$ is trajectory length, and $\Delta_\odot W$ is the increment of $W$ after a complete revolution (Fig. \ref{Fw}a).  The flux along $W$ equals $J=\nu/\Delta_\odot W$. While $P(W)$ is singlevalued, $F(W)$ is multivalued analogous to $W$ with the increment after a complete revolution of $\Delta_\odot F=\alpha \Delta_\odot W$.

Fig. \ref{fig:FW} shows a typical diffusive model describing dynamics in an addev sub-ensemble for a simple Markov chain on a circle. An optimal RC $W$ is a multivalued function which covers the circle analogous to the angle. Diffusive models for more complex, multidimensional systems are essentially the same; they may have a non constant stationary probability and, correspondingly, more complex $F(W)$ defined by Eq. \ref{Fw}. Most complexity is absorbed by the optimal RC $W$, which performs complex mapping from a large multidimensional configuration space of a system of interest onto a circle, analogous to the committor preforming complex mapping from a configuration space onto the $[0,1]$ interval. 

\subsection{Addev master equation}
Consider continuous time Markov chain with rate matrix $K(i|j)$, which defines rate from state $j$ to state $i$. The addev master equation (AME) describes a sub-ensemble of trajectories for which the same RC (phase of the addev) is optimal for both forward and time-reversed dynamics \cite{krivov_addev_2022}. It is a system of equations with three vectors of unknowns $S(i,t)$, $u(i,t)$, and $v(i,t)$
\begin{subequations}
	\label{AME}
	\begin{align}
		\sum_j K(j|i)\frac{u(j,t)}{u(i,t)}[S(j,t)-S(i,t)]&=-\frac{dS(i,t)}{dt}\\
		\sum_j K(i|j)\frac{v(j,t)}{v(i,t)}[S(i,t)-S(j,t)]&=-\frac{dS(i,t)}{dt}\\
		\sum_j K(i|j)\frac{u(i,t)}{u(j,t)}P(j,t)- \sum_j K(&j|i)\frac{u(j,t)}{u(i,t)}P(i,t)\nonumber \\ &=\frac{dP(i,t)}{dt}
	\end{align}
\end{subequations}
where $P(i,t)=u(i,t)v(i,t)$. The AME describes evolution of probability distribution $P$, together with axillary phase function $S$, in a time-reversible manner; it can be integrated forward and backward in time. In this manuscript, however, we focus only on stationary solutions of the AME, which provide optimal RCs.

A stationary solution, $S(i,t)=W(i)-\nu t$, $u(i,t)=u(i)$ and $v(i,t)=v(i)$ describes an addev sub-ensemble of trajectories. We call the tuples $(S, u, v)$ or $(S, R)$ an addev with $S$ or $W$ being the phase of the addev, $u$ and $v$ being the forward and time-reversed biasing factors, respectively, $R(i)=\sqrt{u(i)v(i)}$ being the module or amplitude of the addev, and $\nu$ the addev eigenvalue. The biasing factors $u$ and $v$ define the rate matrix 
\begin{subequations}
	\label{kij}
	\begin{align}
		&\tilde{K}(j|i)=K(j|i)\frac{u(j)}{u(i)} \quad \mathrm{for} \, i \ne j,\\
		&\sum_j \tilde{K}(j|i)=0,
	\end{align}
\end{subequations}
that specifies the stochastic conservative dynamics of the sub-ensemble of trajectories and the corresponding stationary probability $P(i)=u(i)v(i)$; $\tilde{K}(i|i)$ is defined via Eq. \ref{kij}b. For this sub-ensemble of trajectories, $W$ is an optimal RC for both forward and time-reversed dynamics (Eqs. \ref{AME}a and \ref{AME}b), meaning non-equilibrium dynamics of the sub-ensemble can be described by a single RC.  The overall dynamics of trajectories projected on $W$ is rather simple, they move (on average) with constant velocity $\nu$ along the periodic coordinate $W$ (Fig. \ref{fig:FW}). We call such a solution, representing a sub-ensemble of trajectories performing stochastic periodic motion a stochastic eigenmode. The AME defines $S$ or $W$ and $\nu$ up to an overall factor, which can be fixed by specifying the increment of $W$ after a complete revolution $\Delta_\odot W$.

The stochastic dynamics in an addev sub-ensemble is completely specified by the biased rate matrix $\tilde{K}(i|j)$, which is known once an addev solution is found. Thus, knowing $\tilde{K}$ one can, in principle, compute any property of the dynamics, however, some properties can be readily computed just from the addev $(W,u,v)$, or $(W,R)$,  as, for example, the stationary probability. Next, we describe some of such properties.

\subsection{Expressions for committors}
\label{sect:q}
A harmonic function $h(i)$ for a Markov chain with rate matrix $K$ is defined as
\begin{equation}
	\label{q0}
	\sum_j K(j|i)[h(j)-h(i)]=0
\end{equation}
The committor function $q(i)$ can be found from a harmonic function as $q(i)=ah(i)+b$, where constants $a$ and $b$ are determined from the two boundary conditions, for example, for $q_B$ they are $q_B(A)=0$ and $q_B(B)=1$. Such a defined committor makes sense only for nodes between $A$ and $B$, i.e., for nodes with $0\le q\le 1$. We denote the relation $q=ah+b$ as $q\propto h$; any such $q$ is also a solution of Eq. \ref{q0}. 

Assume that the Markov chain satisfies the detailed balance $K(i|j)\pi(j)=K(j|i)\pi(i)$, where $\pi(i)$ is the equilibrium probability. Then Eq. \ref{AME}c can be rewritten as
\begin{equation}
	\label{q1}
	\sum_j K(j|i)\frac{u(j)}{u(i)}\left(\frac{v(j)}{\pi(j)u(j)}-\frac{v(i)}{\pi(i)u(i)}\right)=0.
\end{equation}
Comparing with Eq. \ref{q0}, one sees that $q(j)\propto \frac{v(j)}{\pi(j)u(j)}$ is the committor function for $\tilde{K}(j|i)$ (Eq. \ref{kij}), the conditioned dynamics of the sub-ensemble.
Analogously, for the time-reversed dynamics one can rewrite Eq. \ref{AME}c as
\begin{equation}
	\label{q2}
	\sum_j K(i|j)\frac{v(j)}{v(i)}\left(\frac{\pi(j)u(j)}{v(j)}-\frac{\pi(i)u(i)}{v(i)}\right)=0,
\end{equation}
meaning that committor for the time-reversed dynamics is $q'(j)\propto \frac{\pi(j)u(j)}{v(j)}$.

The diffusive model of dynamics along addev RC $W$ suggests the following dependence of the committor as a function of $W$: $dq/dW\sim e^{\beta F(W)}/D(W)\sim P(W)^{-1}e^{-\alpha W}/D(W)=1/Z_{C,1}(W)e^{-\alpha W}$. Since $Z_{C,1}(W)$ is constant \cite{krivov_addev_2022}, one finds that $q(W)\propto e^{-\alpha W}$. Analogously, for the time-reversed committor one finds $q'(W)\propto e^{\alpha W}$. We next show that for a system with the detailed balance the suggested dependence is correct, that committor computed from the unprojected addev sub-ensemble of trajectories coincides with that predicted by the diffusive model. 

Assume that the detailed balance is satisfied, then difference of Eqs. \ref{AME}a,b equals
\begin{equation}
	\label{q3}
	\sum_j K(j|i)\frac{u(j)}{u(i)}[q(j)+q(i)][W(j)-W(i)]=0
\end{equation}
Considering the obtained equation as a system of linear equations on variables $[q(j)+q(i)][W(j)-W(i)]$, one sees that the system is identical to Eq. \ref{q1} written as $\sum_j K(j|i)u(j)/u(i)[q(j)-q(i)]=0$. Assume that this system has a unique solution, then it means that the two solutions are proportional, i.e., 
$q(j)-q(i)=-\alpha/2 [q(j)+q(i)][W(j)-W(i)]$, where $\alpha$ is some constant. This system of linear equations can be used to find $q$ as a function of $W$. If $\delta q=q(j)-q(i)\ll q(i)$, then $[q(j)-q(i)]/[q(j)+q(i)]=1/2 [\ln q(j)-\ln q(i)] + o (\delta q^2)$ is accurate to the second order, which gives an approximate solution of the system of linear equations as $\ln q(j)=-\alpha W(j)$. In case when Eqs. \ref{q1} and \ref{q3} have more than one solution, we fix the solution by specifying the boundary conditions $q(W_A)=0$ and $q(W_B)$=1, i.e., we are not interested in any committor but in a committor between any  two iso-surfaces of $W.$ Such a solution is obviously unique and it is satisfied by $q(j)\propto e^{-\alpha W(j)}$.  

Summing up Eqs. \ref{AME}a,b and replacing $1-q(j)/q(i)$ with $\alpha/2 [1+q(j)/q(i)][W(j)-W(i)]$ one obtains
\begin{equation}
	\label{q4}
	\alpha \sum_j K(j|i)\frac{u(j)}{u(i)}[1+\frac{q(j)}{q(i)}][W(j)-W(i)]^2/4=\nu
\end{equation}
Assuming $q(j)/q(i)\approx 1$ (a fine discretization provided by a Markov chain), one finds $\alpha=\nu/\langle D \rangle$, where $\langle D \rangle=\sum_{ij} K(j|i)u(j)v(i)[W(j)-W(i)]^2/2$ is the average diffusion coefficient along $W$.

$\langle D \rangle$ can be also computed from a long stationary trajectory as $\langle D \rangle=\int Z_{C,1}(W)dW/T=Z_{C,1}(W)\Delta_\odot W/T$ \cite{krivov_addev_2022}, which leads to Eq. \ref{Fw}b.

Combining the two expressions for the committors one finds $u=Re^{\alpha/2 W}$ and $v=Re^{-\alpha/2 W}$, where $R=\sqrt{P}$ and we assumed $\pi=1$. Since $W$ is multivalued, it follows that for a system with the detailed balance $u$ and $v$ are multivalued too. Section \ref{sect:idf} presents an example of a system without the detailed balance, where $q\propto e^{-\alpha W}$, while $u$ and $v$ are single-valued or even constant.

Consider a cut in a Markov chain projected onto $W$ at particular value of RC, $W_0$. Namely, remove all the transitions/rates between nearby states on the different sides of $W_0$. Such a cut does not perturb the rates between states far from the cut, however, it removes the flux. The equilibrium probability for the rate matrix of the addev sub-ensemble $\tilde {K}$ with such a cut, can be computed as $\tilde{\pi}(i)=\pi(i)u^2 (i)$  (see Eq. \ref{kij}a). The probability is a multivalued function, analogous to $u$; different brunches are obtained by moving the cut along $W$. It can be related to the stationary probability of non-equilibrium dynamics with the flux $P(i)=u(i)v(i)$, as $\tilde{\pi}(i)=P(i)/q(i)=P(i)e^{\alpha W}$, or in terms of the free energy profile of the diffusive model $\beta F(W)=-\ln \tilde{\pi}(W)=-\ln P(W) -\alpha W$, i.e., Eq. \ref{Fw}a. 

Summarising, the diffusive model along the addev optimal RC $W$ can be used to compute exactly the forward and time-reversed committors, and hence the equilibrium flux, the mfpt, and the mtpt between any two points (iso-surfaces) along the RC for the addev sub-ensembles of trajectories. 

\subsection{Expressions for various stationary properties of an addev sub-ensemble}
The equilibrium flux across any surface can be computed as 
\begin{equation}
	\label{J}
	J=\sum_{ji} \mathrm{sign}(j,i) u(j)K(j|i)v(i),
\end{equation}
where function $\mathrm{sign}(j,i)$ is $1$, $-1$ and $0$ if transition $i \rightarrow j$ crosses the surface in the positive direction,  negative direction or does not cross the surface, respectively. 


For systems where $q(W)=e^{-\alpha W}$, given two boundary states $W_A$ and $W_B$ the forward committor function, which satisfies the boundary conditions $q_B(W_B)=1$ and $q_B(W_A)=0$ can be computed as $q_B(W)=[q(W)-q(W_A)]/[q(W_B)-q(W_A)]$. Analogously, the time-reversed committor function $q'_A(W)$, which satisfies the boundary conditions $q'_A(W_A)=1$ and $q'_A(W_B)=0$ can be computed as $q'_A(W)=[q'(W)-q'(W_B)]/[q'(W_A)-q'(W_B)]$, where $q'(W)=e^{\alpha W}$. Knowing the committors one can compute properties of the TPs in the addev sub-ensemble of trajectories. For example, for TPs $A\rightarrow B$, the stationary distribution equals \cite{metzner_transition_2009}
\begin{equation}
	\label{PTP}
P^{TP}(i)=q_B(i)u(i)v(i)q'_A(i)=q_B(i)P(i)q'_A(i);
\end{equation}
the stationary flux is 
\begin{equation}
J^{TP}=\sum_{ji} \mathrm{sign}(j,i) q_B(j)u(j)K(j|i)v(i)q'_A(i);
\end{equation}
the mtpt is $Z^{TP}/J^{TP}$, where $Z^{TP}=\sum_i P^{TP}(i)$ is the total weight of the TP sub-ensemble. Similar expressions can be obtained for TPs $B\rightarrow A$, $A\rightarrow A$, and  $B\rightarrow B$. 

The probability to observe an addev sub-ensemble for time-interval $T$ can be estimated as $e^{-T/\tau A_\tau}$, where $A_\tau=D(\tilde{P}_\tau||P_\tau)=\sum_{ij}\tilde {P}_\tau(i|j)P(j)\ln[\tilde {P}_\tau(i|j)/P_\tau(i|j)]$ is the Kullback-Leibler (KL) divergence of $\tilde{P}_\tau$ from $P_\tau$ and $\tilde{P}_\tau=e^{\tau \tilde{K}}$ and $P_\tau=e^{\tau K}$ are the transition probability matrices \cite{Csiszar_conditional_1987,Touchette_large_2009}. By taking the limit $\tau \rightarrow 0$, the probability to observe an addev can be expressed as $e^{-TA}$, where  
\begin{equation}
A=\sum_{ij: i\ne j} \tilde{K}(j|i)P(i) g[K(j|i)/\tilde{K}(j|i)],
\end{equation}
and $g(x)=-\ln x -1 +x$. For an addev biased rate matrix $x=u(i)/u(j)$ and assuming $x-1 \ll 1$, one obtains $g(x)\approx (x-1)^2/2$. For a Markov chain with the detailed balance with $\pi(i)=const$ and an addev with $P(i)=const$,  one finds that $u(i)\sim e^{\alpha/2 W(i)}$, $g(u(i)/u(j))\approx \alpha^2/8 [W(i)-W(j)]^2$, and $A=\alpha^2 \langle D \rangle/4=\alpha \nu/4$. 

\subsection{Gauge invariance}
\label{sect:gauge}
Let $(S(i),u(i),v(i))$ be an addev for the rate matrix $K(i|j)$. It is straightforward to verify that $(S(i),u(i)/\gamma(i),v(i)\gamma(i))$  is an addev for the transformed rate matrix $K(i|j)\gamma(i)/\gamma(j)$, i.e., they satisfy the AME. Note, that this transformation keeps invariant the observable properties of the addev $\tilde{K}(j|i)$, e.g.,  $P(i)$, $S(i)$, $q(i)$ and $q'(i)$. For example, for the committors, it follows from the fact that equilibrium probability $\pi$ is transformed as $\pi(i)\gamma^2(i)$. Note that $u(i)$ and $v(i)$, which are changed by the transformation can not be observed directly. We call this invariance a gauge invariance. It has the following consequences. First, the standard way to introduce a non-uniform equilibrium probability or a potential into the stochastic dynamics is to modify the rates by $\gamma(i)=e^{-\beta F(i)/2}$. The additive eigenvectors are invariant to such a way of introducing the potential, which mean one needs to introduce the potential into the addev framework in a different way, we describe the proper way in Section \ref{sect:potential}. Second, a conveniently chosen gauge may simplify the derivation of equations.

\subsection{Sampling an addev sub-ensemble by solving the AMU ``on the fly''}
Stochastic dynamics in the addev sub-ensemble is described by the biased rate matrix $\tilde{K}$ (Eq.
\ref{kij}), which is known, once an addev solution is found. Evolution of probability distribution is described by Eq. \ref{AME}c, where however, $P$ is now not the stationary probability of the addev sub-ensemble $u(i)v(i)$, but arbitrary. Alternatively, one can compute a stochastic trajectory of the addev sub-ensemble directly, by simulating stochastic dynamics of Markov chain with $\tilde{K}$. For Markov chains with regular configuration space, e.g., a lattice in n-dimensional Euclidean space, considered below, such a trajectory, which samples a stationary addev solution, can be computed by solving the AMU equation ``on the fly''. Define the number of degrees of freedom (d.f.) of an addev equation/solution as the smallest number of variables that uniquely specify the solution. Note that without the loss of generality one can set $u(i_0)=1$, $v(i_0)=1$, $W(i_0)=0$ for some initial $i_0$.  Specifying these d.f. at one point, one is able to compute their values at any other point by propagating the solution of the equation along particle's trajectory. The solution defines the local biased rate matrix $\tilde{K}$, which can be used to propagate the system's trajectory, which allows an efficient sampling scheme of an addev sub-ensemble; one does not need to solve an addev equation in the entire configuration space beforehand; which is especially useful in a high-dimensional configuration space. It also can help simplify analysis of multivalued addev solutions. Instead of cumbersome specification of a global addev solution, one may just follow a local solution along system's trajectory. To find an addev solution that satisfies specific boundary conditions, a shooting method can be used. 

\section{Analysis of diffusion. No internal degrees of freedom.}
\label{sect:classical}
\subsection{Derivation of the PDE} We start with a Markov chain describing one-dimensional diffusion (with position dependent diffusion coefficient) with the following rate matrix $K(i\pm 1|i)=\frac{D(i)+D(i\pm1)}{2\Delta x^2}$ \cite{bicout_electron_1998}, here $D(i)$ is the diffusion coefficient and $\Delta x$ is the distance between the states of the Markov chain. The Markov chain satisfies the detailed balance with constant equilibrium probability $\pi(i)$. For the general case of diffusion in a potential the rate contains factor $e^{-\beta/2 (U(i\pm1)-U(i))}$, which however can be removed by a gauge transformation (section \ref{sect:gauge}). The AME for the Markov chain is

\begin{widetext}
\begin{subequations}
	\label{AMEd1}
	\begin{align}
K(i+1|i)\frac{u(i+1)}{u(i)}[S(i+1)-S(i)]+K(i-1|i)\frac{u(i-1)}{u(i)}[S(i-1)-S(i)]=&-\frac{dS(i)}{dt}\\
K(i|i+1)\frac{v(i+1)}{v(i)}[S(i)-S(i+1)]+K(i|i-1)\frac{v(i-1)}{v(i)}[S(i)-S(i-1)]=&-\frac{dS(i)}{dt}\\
K(i|i-1)\frac{u(i)}{u(i-1)}P(i-1)+K(i|i+1)\frac{u(i)}{u(i+1)}P(i+1)  -&  \nonumber \\  K(i+1|i)\frac{u(i+1)}{u(i)}P(i)-K(i-1|i)\frac{u(i-1)}{u(i)}P(i)=&\frac{dP(i)}{dt}.
	\end{align}
\end{subequations}
\end{widetext}
Expanding to the second order of $\Delta x$ and taking the limit of $\Delta x\rightarrow 0$, one arrives at the following system of PDE

\begin{subequations}
	\label{addev1}
	\begin{align}
	(DPS')'=& 0\\
	-DS'(\ln q)'= & -\dot{S}\\
	(DP(\ln q)')'= & \dot{P},
	\end{align}
\end{subequations}
where dot and prime denote partial derivatives $\partial/\partial t$ and $\partial/\partial x$, respectively, and $q=v/u$.  For a stationary addev solution, $\dot{P}=0$, by comparing Eqs \ref{addev1} a and c, one has $(\ln q)'=-\alpha S' $ in agreement with section \ref{sect:q}. Since $S$ is defined up to an overall constant, one convenient normalisation is  $\alpha=1$, which leads to the equation on $S$
\begin{equation}
	\label{addev3}
	D(S')^2=-\dot{S}   
\end{equation}
The equation has the form of the classical Hamilton-Jacobi equation for the action function $S$, where $D\sim 1/m$ and $-\dot{S}=\nu\sim E$. From Eq. \ref{addev1}a or c one obtains $P\sim 1/(DS')$, which corresponds to the classical $P\sim 1/v$, where $v\sim DS'\sim p/m$ is velocity and $p\sim S'$ is momentum. Note that S is dimensionless while $\nu$ has the dimension of rate, they differ from their classical counterparts, action $S$ and energy $E$, by an overall factor with dimension of action; a convenient choice is the Planck constant $\hbar$, which, by the special choice of units can be made equal to $1$ and which is assumed henceforth. The Planck constant is taken here for convenience, one can use any other constant factor with dimension of action, which would amount to the change of measuring units or the overall scale factor of $S$. Extension to higher dimensions is presented in appendix.

\subsection{Addev description of diffusion on the line}
\label{diffusion_on_the_line}
Consider diffusion on the line with constant diffusion coefficient $D$. Solution of Eq. \ref{addev3} is $S(x,t)=kx-\nu t$, where $\nu=k^2D$ and $k$ is arbitrary. The addev describes a sub-ensemble of trajectories performing biased diffusion. As an optimal RC one can use either $W$ or $x$ as they are related one-to-one. Consider first $x$ as the RC. The diffusive model describing the addev dynamics has $D$ as diffusion coefficient and $\beta F(x)=-\alpha W=-kx$ as free energy profile. The model describes simple dynamics of the trajectories moving/drifting at constant speed of $v=Dk$. Its classical analogy is a particle moving with constant speed $v$ on the line. Consider $W$ as the RC. The addev trajectories drift at constant speed of $\nu$ along $W$ (by definition); the two speeds are related by the Jacobian $\partial W/\partial x=k=\nu/v$. Its classical analogy is the movement of the point depicting the system along the action function with velocity $E$. For the committors one obtains $q\propto e^{-kx}$ and $q'\propto e^{kx}$, which allows one to describe TPs in the addev sub-ensembles, as was done in Fig. \ref{fig:TPtau}, using Eq. \ref{PTP}.

The probability to observe such an addev for time interval $\Delta t$ can be computed using the KL divergence as $P\sim e^{-\Delta t Dk^2/4}=e^{-\Delta t \nu/4}=e^{-\Delta t \frac{v^2}{4D}}=e^{-\frac{\Delta x^2}{4D\Delta t}}$, here $\Delta x=v\Delta t$. The fact that it equals $e^{-\Delta t \nu/4}$ follows from constant probabilities $\pi$ and $P$ and the detailed balance. In this case the addev sub-ensemble can be obtained by standard conditioning of diffusion trajectories on displacement $\Delta x$ after time interval $\Delta t$ and the probability to observe such a sub-ensemble is the standard probability density of finding a particle around $\Delta x$ after time interval $\Delta t$.

Consider now a long diffusion trajectory that visits a sequence of addevs. Assuming that addevs are independent, the probability to observe such a sequence takes a suggestive form $\sim \exp( -\sum_i \frac{\Delta x_i^2}{4D\Delta t_i})$. Summing over all possible sequences of addevs with the corresponding probability one arrives at expression analogous to the standard path summation/integral for diffusion. An important, subtle difference here is that the described summation is over addevs and each addev describes a sub-ensemble of trajectories not a single trajectory as in the standard path integral. Correspondingly, a particular trajectory has a non-zero probability to appear in many such sequences/paths of sub-ensembles. 

\subsection{Addev description of diffusion on the interval}
\label{sect:diff_segm}
Now consider diffusion on the interval $[0,L]$ with constant diffusion coefficient $D$. Here the system repeatedly moves from $0$ to $L$ and backward, which can be considered as a cartoon model of a reversible reaction, e.g., protein folding, where a protein folds and unfolds repeatedly. Standard description with committors describes only the TP segments of trajectories, i.e., when the protein either folds or unfolds. The addevs can be used to describe continuous dynamics of repeated folding/unfolding. Solution of Eq. \ref{addev3} on the interval is $S(x,t)=W(x)-\nu t$, where $W'=\pm k$. $W(x)$ is a multivalued function consisting of following branches, connected at the boundaries, $W_{2i}(x)=2ikL+kx$ and $W_{2i+1}(x)=2(i+1)kL-kx$. It has a vertical zig-zag or triangle wave pattern. The addev describes a sub-ensemble of trajectories performing biased diffusion with a constant drift at speed of $Dk$ towards one boundary, then in the opposite direction and so on. Its classical analogue is a particle performing periodic motion with constant speed $v=Dk$ in a square well of width $L$. In principle, an addev with any $k$ can be observed, however if one would like to select a single addev that best describes the diffusion dynamics on the interval, that will be an addev with the same flux or the same mean first passage full-trip time, i.e., $2L/v=L^2/D$ or $k=2/L$. 

Considering the complexity of biasing provided by such addevs, e.g., a bias that switches direction, one may wonder whether such a bias does exist and can be observed in practice. Since addevs describe sub-ensembles of trajectories it is sufficient to specify how the corresponding sub-ensemble of trajectories can be selected. For example, consider all such trajectories that start from some point at distance $\Delta x$ from the boundary, visit the boundary and arrive at the same point after time interval $2\Delta t$. Such a sub-ensemble is certainly not empty. Trajectories in the sub-ensemble perform biased random walk with average velocity of $\Delta x/\Delta t$ initially towards the boundary and then from it.  

A subtle question in modelling such a dynamics of an addev, is when to switch the bias from one direction to the opposite. It should be done when the system reaches a boundary state, however, since the dynamics is stochastic, the system will visit a boundary state many times while it is nearby. A simple solution for this system is to model movement in two different directions by considering two different pathways explicitly. Specifically, forward pathway from A to B and backward pathway from B to A are joined at the corresponding boundary states thus forming a circle with circumference of $2L$. Correspondingly, the first half of the circle describes forward pathway, while the second half describes the backward pathway. The constant bias applied along the entire circle. The system performs biased stochastic periodic motion along the circle and no switch of bias is required. The diffusive model is analogous to that shown on Fig. \ref{fig:FW}. This construction will be used in some examples below.

\section{Addev master equation with potential}
\label{sect:potential}
So far we have used the addevs to describe stochastic dynamics of free diffusion represented by a Markov chain. In particular, we have shown that this dynamics can be decomposed on addevs, or approximated by a path integral over addevs. However many stochastic process are modelled as stochastic dynamics/diffusion in the field of an external force or in a potential. How one should introduce potential in the addev formalism? The straightforward approach could be to use a Markov chain for diffusion with potential, which for a one-dimensional case is $K(i\pm 1 |i)=re^{-\beta/2 [U(i\pm 1) -U(i)]}$. Such a Markov chain has  canonical distribution $\pi(i)\sim e^{-\beta U(i)}$ as the equilibrium probability. Next, one could compute the addevs for this Markov chain and probabilities to observe them. However, as described in section \ref{sect:gauge}, potential, introduced in such a way, does not change addev properties because of the gauge invariance; addev sub-ensembles are exactly as those for $U=0$. Thus, the potential needs to be introduced in a different way.

A guiding principle could be the requirement that description of stochastic dynamics with addevs reproduces the canonical probability distribution $e^{-\beta U(i)}$ of finding the system in state $i$. An important difference with the standard description of stochastic dynamics is that in the latter only a single stationary/equilibrium eigenvector of the standard master equation, that with zero eigenvalue, contributes to the equilibrium probability, while in the addev description all the addevs contribute. It means that the canonical distribution should be obtained as a sum of stationary distributions of all addevs with the corresponding weights. The similarity of the addev equation for diffusion (Eq. \ref{addev3}) with that of the classical mechanics suggests a possibility of introducing the potential in analogy with classical mechanics, so that Eq. \ref{addev3} reads now $D(S')^2+U=-\dot{S}$. This can be done if the AME takes the potential in to account as
 \begin{subequations}
	\label{AMEU}
	\begin{align}
		\sum_j K(j|i)\frac{u(j,t)}{u(i,t)}[S(j,t)-S(i,t)]+U(i)=&-\frac{dS(i,t)}{dt}\\
		\sum_j K(i|j)\frac{v(j,t)}{v(i,t)}[S(i,t)-S(j,t)]+U(i)=&-\frac{dS(i,t)}{dt}\\
		\sum_j K(i|j)\frac{u(i,t)}{u(j,t)}P(j,t)- \sum_j K(j|i)\frac{u(j,t)}{u(i,t)}&P(i,t)\nonumber \\ =\frac{dP(i,t)}{dt}&
	\end{align}
\end{subequations}
It is equivalent to replacing $\frac{dS(i,t)}{dt}$ with $\frac{dS(i,t)}{dt}+U(i)$.
Such an introduction of the potential into the AME preserves the formalism developed so far, in particular that an addev provides an optimal RC. For example, the relationship $q\propto e^{-\alpha W}$ for systems with the detailed balance, derived in section \ref{sect:q}, still holds, because the derivation uses the difference between Eqs. \ref{AMEU}a,b, where $U$ cancels. 
The sum of Eqs. \ref{AMEU} a and b describes the conservation of energy, which for the systems with the detailed balance takes the form (cf. Eq. \ref{q4})
\begin{equation}
	\label{q5}
	\alpha \sum_j K(j|i)\frac{u(j)}{u(i)}[1+\frac{q(j)}{q(i)}][W(j)-W(i)]^2/4+U(i)=\nu,
\end{equation}
where the first term is interpreted as the kinetic energy. From Eq. \ref{q5} it follows that $\alpha \langle D \rangle=\nu - \langle U \rangle$ and from Eq. \ref{AMEU} that the drift speed along $W$ is no longer constant. Note that both $\nu$ and $U$ have the dimension of frequency and to relate them to their classical counterparts, one needs to multiply them by a constant with the dimension of action, for which one can take the Planck constant $\hbar$, which is assumed to be equal 1. Correspondingly, in analogy with classical statistical mechanics, each addev contributes with the classical canonical weight of $e^{-\beta \nu}$.

The canonical weight to observe an addev can be justified as follows. Originally, the addevs were introduced/interpreted as sub-ensembles of trajectories generated by a Markov chain, which allowed us to compute the probability to observe an addev using the KL divergence. However, the addevs and Markov chains introduce potential in different ways, meaning that addevs describe sub-ensembles of trajectories of a stochastic process different from that described by the Markov chains. One can probably relate the two processes by relating the two potentials, analogous the way the eigenfunctions of the diffusion equation are related to that of the Schr\"odinger equation \cite{risken_fokker-planck_1984}, however we do not develop this idea here. We assume that the AME describes a different process and probability to observe an addev should be computed differently. We consider the AME with potential as a general tool to describe/approximate the generic stochastic dynamics, in particular that produced by classical equations of motions or molecular dynamics. Then the properties of the addev description are defined by the dynamics of interest, for example, one may adapt the addev properties to better describe the dynamics of interest. 

Consider systems with classical dynamics, e.g., molecular dynamics simulations. Namely, consider a classical system of interest in contact with a large classical system playing the role of a thermostat. The dynamics of the entire system at fixed energy is described by an addev. Since the classical system can exist indefinitely, we assume that the corresponding addev is a genuine stationary solution of the AME (Eq. \ref{AMEU}), which exists indefinitely, rather than a rare event observed with exponentially decreasing probability of $e^{-AT}$. Correspondingly, according to the classical statistical mechanics arguments the energy of the system of interest, which is in contact with the thermostat, is canonically distributed. It means that the system of interest is described by a collection of addevs, canonically distributed according to their energy/frequency/eigenvalue $\nu$. In practice, one may compute all addevs for an isolated system and sum them up with the corresponding canonical weights, or explicitly consider a system coupled to a thermostat, which however complicates the description. The AME with potential accurately describes the classical dynamics in full configuration space. It suggests that stochastic dynamics, obtained by coarse-graining the classical dynamics can be also approximated by the addev formalism, and probably more accurately than the standard approaches.

Eq. \ref{AMEU} can also be given the following self-consistent interpretation within the addev framework. Assume that there is no potential at the fundamental level of the exact description of stochastic dynamics of a system and Eq. \ref{AME} is exact. However, such a system may have dynamics on two very different time-scales, fast and slow, and we are interested in a coarse-grained  simplified description of the slow dynamics. For exact description of dynamics by Eq. \ref{AME}, it is correct to assume that the change of the phase $S$ happens only when the system changes states and it is described by $W(j)-W(i)$. In the coarse-grained description, while the system stays in the same macrostate, $S$ can still change due to the micro dynamics or the change of microstates. $U(i)$ is the source term which describes the rate of change of $S$, namely $dS/dt$, while the system stays in macrostate $i$, namely (cf. Eq. \ref{S0})
$$S(i,t)-S(i,t+\tau)=\sum_j \tilde{P}_\tau(j|i) [W(j)-W(i)]+ U(i) \tau, $$
here we assumed $\tilde{P}_\tau (i|i)\approx 1$ because $\tau$ is small. Taking limit $\tau\rightarrow 0$ one obtains Eq. \ref{AMEU}a. It is important that $W$ is a multivalued function, which allows it to produce a stationary change of $S$ while being confined to a finite set of microstates in a single macrostate. A simple example of such a coarse-graining is presented below.

\subsection{Addev description of diffusion with potential}
Consider, as an illustrative example, the addev description of stochastic dynamics in a (harmonic) well. The description is a little more complicated than presented in section \ref{sect:diff_segm}. An addev describes a sub-ensemble of trajectories performing biased diffusion with a position dependent drift speed towards a turning point, then in the opposite direction and so on. The position dependent drift speed equals $DW'=\sqrt{D(\nu-U)}$. The stationary probability to observe a system in an addev at position $x$ is proportional to $P(x)\sim 1/(DW')\sim 1/\sqrt{D(\nu-U)}$. The diffusive model is similar to that shown on Fig. \ref{fig:FW}, however, with a non constant stationary probability. Note that higher potential corresponds to higher stationary probability, in contrast to the canonical distribution. Combining all the addevs with canonical weights $e^{-\beta \nu }$, one finds the probability to observe the system at position $x$ equal to $P(x)\sim e^{-\beta U(x)}$. 

Consider next, as a model of a reaction, one-dimensional diffusion on a potential consisting of two minima separated by a barrier of height $U^\dag$. Addevs, with energy higher than the barrier height $\nu>U^\dag$, contribute to the reaction between the two minima. They describe sub-ensembles of trajectories performing biased diffusion with position dependent drift speed from one minimum (turning point) to another, then in the opposite direction and so on.  Each addev contributes with the canonical weight of $e^{-\beta \nu}$, which can be used to compute the properties of the reaction sub-ensemble of trajectories, i.e., the flux, the stationary probability, etc. For small temperatures the main contribution comes from the addevs with energy just above that of the barrier. Addevs with energies less than that of the barrier describe dynamics inside the minima.

In principle, we have obtained the desired description of stochastic reaction dynamics, as a periodic stochastic process, where the system goes from one minimum to the other and back, repeatedly. The description, however, has an obvious shortcoming - it is rather too "deterministic". A defining characteristics of a stochastic process is a possibility of fluctuations, where a trajectory with energy lower than that of a barrier, can gain energy as a result of such a fluctuation and overcome the barrier. In the one-dimensional system considered above, the addevs with energy less than that of the barrier do not overcome the barrier and describe the dynamics inside the minima. This shortcoming might be critical to low dimensional systems only.
In high-dimensional systems with many degrees of freedom, the redistribution of energy between different degrees effectively leads to fluctuations of energy along an optimal RC ($W$). One way to overcome the shortcoming for a low dimensional system is to explicitly couple it to a thermostat, which however, complicates the description. An alternative is described in section \ref{sect:idf}, where we consider an addev with an internal degree of freedom.

\subsection{Stochastic dynamics in the addev sub-ensemble}
\label{sect:pde}
Stochastic dynamics in the addev sub-ensemble is completely characterized by the biased rate matrix $\tilde{K}$ (Eq. \ref{kij}). Evolution of probability distribution is described by Eq. \ref{AMEd1}c (Eq. \ref{AMEU}c in general case), where however, $P$ is now not the stationary probability of the addev sub-ensemble $P^{st}(i)=u(i)v(i)$, but arbitrary. The equation in the limit of $\Delta x\rightarrow 0$ takes the form of diffusion PDE
\begin{subequations}
	\label{pdeaddev1d}
	\begin{align}
	\dot{P}=&-J' \\
	J=&-DP' + DP\ln'(u^2),
	\end{align}
\end{subequations}
where the dot and prime denote time and space partial derivatives, respectively; $J$ is the flux, $u^2=P^{st}/q$, $-(\ln q)'=W'$ and $P^{st}\sim (D(\nu-U))^{-1/2}$. The stationary (long time limit) solution has stationary probability reproducing that of classical mechanics $P=P^{st}$ and constant but non-zero flux $J=-DP^{st}(\ln q)'=DP^{st}W'$. The equation is an interesting counterpart of the Smoluchowski equation, which, at equilibrium, has canonical probability distribution and zero flux \cite{risken_fokker-planck_1984}. 

Such diffusion can be efficiently simulated/sampled by computing trajectory using time-discretized SDE 
\begin{equation}
	\label{lanj}
	x(t+\Delta t)=x(t)+ D\ln'(u^2)\Delta t +\sqrt{2D \Delta t} \xi,
\end{equation}
where $\Delta t$ is integration time step, $\xi$ is a normally distributed random variable with zero mean and unit variance and $D$ is assumed to be constant; $\ln'(u^2)= \left( W'+\frac{U'}{2(\nu-U)}\right)$,  $W'$ is determined/updated from energy conservation $DW'^2+U=\nu$. Note that dynamics here is entirely in the configuration space; an effective change of the momentum is due to change into a position with different gradient $W'$. Extension to higher dimensions is described in the appendix. By sampling the canonical distribution for $\nu$ one should be able to sample the canonical ensemble. Here, since the addev dynamics is described by PDEs (Eqs. \ref{addev1} and \ref{pdeaddev1d}) it is more convenient to sample dynamics using the SDE; an example of solving the AME (Eq. \ref{AMEU}) ``on the fly'' is provided in section \ref{sect:fly}.

\begin{figure}[htbp]
	\centering
	\includegraphics[width=.9\linewidth]{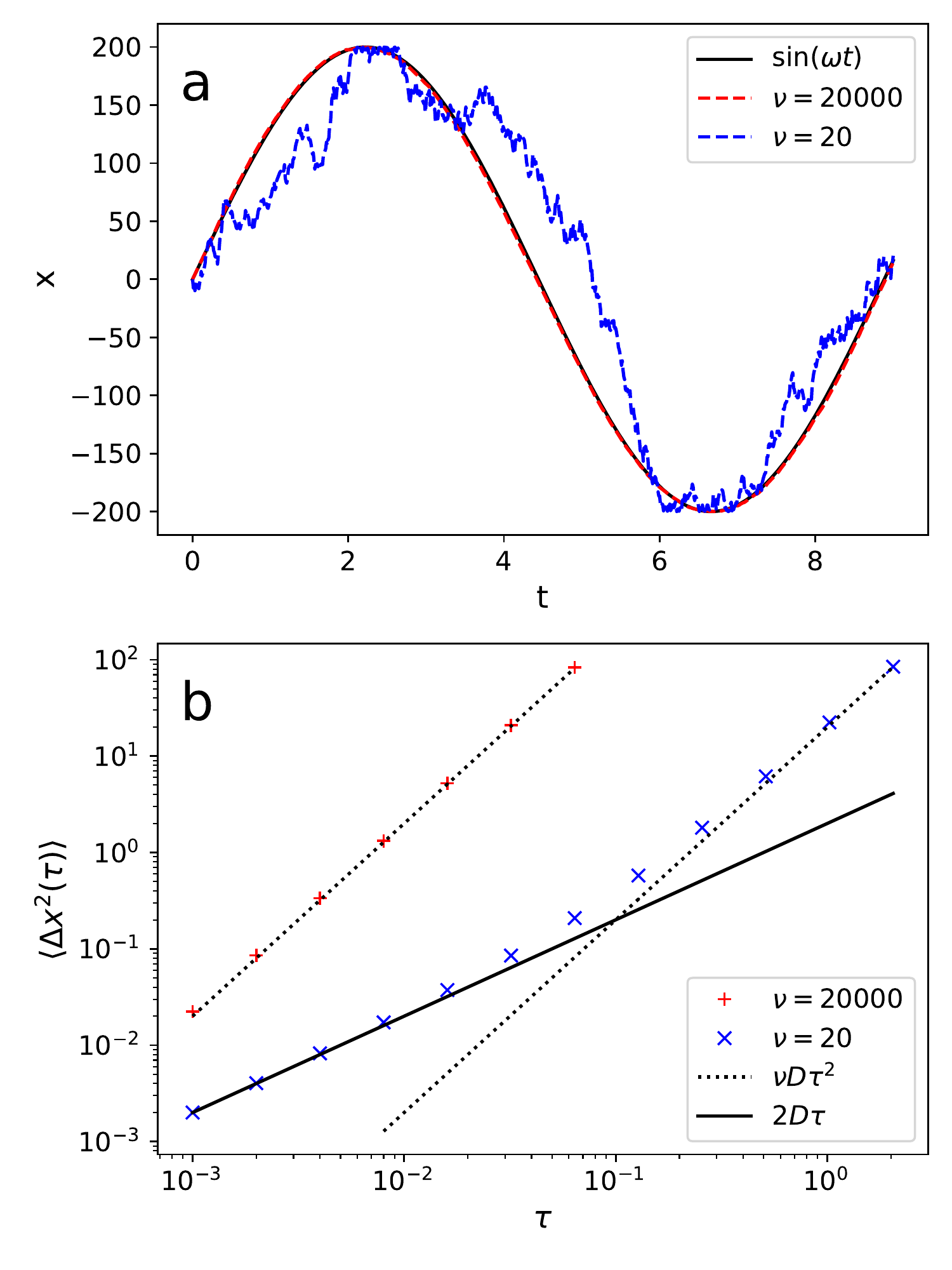}
	\caption{Integration of the SDE (Eq. \ref{lanj}). Panel a) show trajectories for harmonic potential $U(x)=x^2/2$. Black line shows analytical classical solution $\sqrt{2\nu}\sin(\omega t)$, where $\omega =\sqrt{D/2}$. Red and blue dashed lines show stochastic trajectories with eigenvalue/energy $\nu=20000$ and $\nu=20$, respectively; the latter trajectory is scaled up by the factor of $10^{3/2}$.  Panel b) shows $\langle \Delta x^2(\tau) \rangle$ for system with $U=0$. Red pluses and blue crosses show that for trajectories with $\nu=20000$ and $\nu=20$, respectively; black solid and dotted lines show diffusive $2D\tau$ and ballistic $\nu D \tau^2$ behaviour, respectively. For details see text.}
	\label{fig:osc}
\end{figure}

Fig. \ref{fig:osc}a shows stochastic trajectories obtained by integrating the SDE for a system with harmonic potential $U(x)=x^2/2$ and $D=1$. The trajectory with large energy/momentum is in very good agreement with a classical solution, while that with low energy/momentum exhibit stochastic motion. Both trajectories are diffusive at short time-scales, while become ballistic at longer time-scales. To illustrate this point, we compute trajectories with the same energies for a system without the potential. Fig. \ref{fig:osc}b shows the mean squared displacement as a function of time $\langle \Delta x^2(\tau) \rangle=\langle [x(t+\tau)-x(t)]^2 \rangle$ for both trajectories. They are described by $2D\tau$ and $v^2\tau^2\sim \nu D\tau$ in the diffusive and ballistic regimes, respectively. The transition from the diffusive to ballistic regime happens at $v^2\tau^2 \sim 2D\tau$. This time-scale is very small for the trajectory with $\nu=20000$, which is why it is well described by a classical solution. For trajectory with $\nu=20$, the ballistic approximation starts to become accurate for $\Delta x>3$, which is rather close to the size of the accessible configuration space of $2\sqrt{2\nu}\sim 13$.

As described in introduction, one motivation of the developing the addev framework is to obtain an optimal RC for the description of non-equilibrium classical dynamics in the phase space. The described family of addev solutions for diffusion provides a model of classical dynamics; i.e., it reproduces the equilibrium flow and the probability distribution. Thus, instead of considering a classical system of interest, one can consider the corresponding addev, where $W$ is the optimal RC, which provides a solution to the problem. A classical system is translated to the addev diffusive model by mapping the classical $S/\hbar$, $E/\hbar$, $U/\hbar$, $\hbar/m$ to addev's $S$, $\nu$, $U$, $D$. The criterion of the validity of the ballistic description $D\tau\ll v^2\tau^2$ takes the form  $D/(\Delta x v)=\lambdabar/\Delta x\ll 1$, where $2\pi \lambdabar$ is the de Broglie wavelength. 

\section{Analysis of diffusion. Addevs with an internal degree of freedom.}
\label{sect:idf}
We next consider a different addev description of the same Markov chain for one-dimensional diffusion \cite{krivov_addev_2022}. We introduce an internal degree of freedom, which leads to a different expression for the addevs and for the optimal RC. It, correspondingly, describes different sub-ensembles of conditioned trajectories of the same diffusion process.

One dimensional random walk, if the spatial dependence is neglected, can be considered as two-state
dynamics, where the system moves right or left and switches between the two such states with rate $r$. We introduce an internal degree of freedom, describing in which direction the system currently moves; it equals $1$ or $2$ when the systems moves right or left, respectively. The reaction rate matrix in the extended configuration space is $K(i+1,1|i,1)=K(i-1,2|i,2)=K(i+1,1|i,2)=K(i-1,2|i,1)=r$, where $r=D/\Delta x^2$. For simplicity we consider constant diffusion coefficient, which can always be done for one-dimensional diffusion by proper rescaling/transformation of the coordinate.

Note, that while the original Markov chain dynamics in the configuration space is equilibrium, dynamics in the extended configuration space is not, it does not satisfy the detailed balance, as there are no reverse rates $K(i-1,1|i,1)$ and $K(i+1,2|i,2)$. Hence, the constant $\pi(i)$ is the stationary but not equilibrium probability.

AME (Eq. \ref{AME}) for the rate matrix in the extended configuration space reads (the internal degree of freedom is represented by a subscript while the dependence on time variable is  omitted for brevity)
\begin{widetext}
	\begin{subequations}
		\label{AMEdIDF}
		\begin{align}
		r\frac{u_1(i+1)}{u_1(i)}[S_1(i+1)-S_1(i)]+r\frac{u_2(i-1)}{u_1(i)}[1+S_2(i-1)-S_1(i)]=-\frac{dS_1(i)}{dt}\\
		r\frac{u_2(i-1)}{u_2(i)}[S_2(i-1)-S_2(i)]+r\frac{u_1(i+1)}{u_2(i)}[1+S_1(i+1)-S_2(i)]=-\frac{dS_2(i)}{dt}\\
		r\frac{v_1(i-1)}{v_1(i)}[S_1(i)-S_1(i-1)]+r\frac{v_2(i-1)}{v_1(i)}[1+S_1(i)-S_2(i-1)]=-\frac{dS_1(i)}{dt}\\
		r\frac{v_2(i+1)}{v_2(i)}[S_2(i)-S_2(i+1)]+r\frac{v_1(i+1)}{v_2(i)}[1+S_2(i)-S_1(i+1)]=-\frac{dS_2(i)}{dt}\\
		ru_1(i)v_1(i-1)+ru_1(i)v_2(i-1)- ru_1(i+1)v_1(i)-ru_2(i-1)v_1(i)=\frac{d[u_1(i)v_1(i)]}{dt}\\
		ru_2(i)v_2(i+1)+ru_2(i)v_1(i+1)- ru_1(i+1)v_2(i)-ru_2(i-1)v_2(i)=\frac{d[u_2(i)v_2(i)]}{dt}
		\end{align}
	\end{subequations}
\end{widetext}
Here, the complex multivalued addev phase function $W^\mathrm{mv}$, and, correspondingly, $S^\mathrm{mv}=W^\mathrm{mv}-\nu t$, is represented as $W^\mathrm{mv}_j(i)=d^\mathrm{mv}_j+W_j(i)$, where $W_j(i)$ is a singlevalued function, while $d^\mathrm{mv}_j$ is a multivalued simple constant function such that $d^\mathrm{mv}_1-d^\mathrm{mv}_2=d^\mathrm{mv}_2-d^\mathrm{mv}_1=1$; where we used convenient normalization $\Delta_\odot W^\mathrm{mv}=\Delta_\odot d^\mathrm{mv}=2$ \cite{krivov_addev_2022}. The equation has very diverse families of stationary solutions, addevs, with interesting properties, which we next illustrate. In particular, we show how they can be used to describe the stochastic dynamics.

\subsection{Plane wave solutions}
Consider first a simple family of plane wave solutions $u_1(i,t)=v_1(i,t)=\sqrt{\nu+ck}$, $u_2(i,t)=v_2(i,t)=\sqrt{\nu-ck}$, $S_1(i,t)=S_2(i-1,t)=i\Delta x k-\nu t$; where $c=r\Delta x$, and $\nu$ and $k$ are related by the dispersion relation $\nu^2=r^2+k^2c^2$. A stochastic trajectory from this sub-ensemble performs a biased random walk, with a non-zero mean drift speed of $v=r \Delta x (P_1-P_2)/(P_1+P_2)=c^2k/\nu$. Expressing $\nu$ as a function of easily observable $v$ leads to familiar relativistic expression  $\nu=r/\sqrt{1-v^2/c^2}$ \cite{krivov_addev_2022}. The probability to observe such an addev, computed using KL divergence is $\sim e^{-L(v)\Delta t}$, where $L(v)=r(r/\nu-1)=r\sqrt{1-v^2/c^2}-r$, which for small $v$ gives $e^{-\frac{\Delta x^2}{2D\Delta t}}$, i.e., similar to that in section \ref{diffusion_on_the_line}, however with twice the exponent and the maximal speed bounded by $\pm c$, which is due to different biasing. Also, since $r=D/\Delta x^2$ one can not take directly the limit of $\Delta x=0$, instead one considers a very small but finite $\Delta x$. One can however, develop a "non-relativistic" approximation, where such a limit can be taken.

This solution can be also used to illustrate that for systems without the detailed balance one does not have simple relations $q\propto v/(u\pi)$ or $q \propto e^{-\alpha W}$. The committor functions for these solutions are $q_1(i)=q_2(i)\propto (u_2/u_1)^{i}$. However, here, $v/(u\pi)$ is constant. The committors can not be expressed as $q_i\propto e^{-\alpha W^\textrm{mv}_i}$ nor exactly as 
$q_i\propto e^{-\alpha W_i}$, since $q_1(i)=q_2(i)$, while $W_1(i)=W_2(i-1)$. However, in the limiting case of
$\Delta x \rightarrow 0$, where $W_1\approx W_2$, one can derive that $q\propto e^{- W}$, as we show below. 

\subsection{Partial differential equations (PDE)}
A large family of solutions of Eq. \ref{AMEdIDF} can be accurately approximated by the corresponding solutions of the PDE derived below. The PDE in turn can be approximated by the one-dimensional Dirac and Schr\"odinger PDEs, which allows one to develop an intuition about the generic properties of Eq. \ref{AMEdIDF} solutions. The PDEs are derived for general, non-stationary case.

Assume that functions $u$, $v$ and $S$ are sufficiently smooth so that the finite differences can be approximated by the corresponding derivatives. 
Let $u_j(i,t)=R_j(i,t)\alpha_j(i,t)$, $v_j(i,t)=R_j(i,t)/\alpha_j(i,t)$; then $\alpha_j$ can be expressed as functions of $S_j$ using Eqs. \ref{AMEdIDF}a-c and Eq. \ref{AMEdIDF} can be simplified to
\begin{subequations}
	\label{diff:dirl}
	\begin{align}
		\frac{\partial S_1}{\partial t}+c \frac{\partial S_1}{\partial x} &=-r\frac{R_2}{R_1}\sqrt{1-(S_1-S_2)^2}\\
		\frac{\partial S_2}{\partial t}-c \frac{\partial S_2}{\partial x} &=-r\frac{R_1}{R_2}\sqrt{1-(S_1-S_2)^2}\\
		\frac{\partial R_1}{\partial t}+c \frac{\partial R_1}{\partial x}&=rR_2\frac{S_2-S_1}{\sqrt{1-(S_1-S_2)^2}}\\
		\frac{\partial R_2}{\partial t}-c \frac{\partial R_2}{\partial x}&=rR_1\frac{S_1-S_2}{\sqrt{1-(S_1-S_2)^2}}
	\end{align}
\end{subequations}
Consider the one-dimensional relativistic Dirac equation ($r=mc^2/\hbar$)
\begin{subequations}
	\label{diff:dirac}
	\begin{align}
		\frac{\partial \psi_1}{\partial t}+c \frac{\partial \psi_1}{\partial x} =-ir\psi_2\\
		\frac{\partial \psi_2}{\partial t}-c \frac{\partial \psi_2}{\partial x} =-ir\psi_1
	\end{align}
\end{subequations}
Using $\psi_j=R_je^{iS_j}$ one obtains equations on $R_j$ and $S_j$ (notice that $S_j$ are dimensionless)
\begin{subequations}
	\label{diff:dirac2}
	\begin{align}
		\frac{\partial S_1}{\partial t}+c \frac{\partial S_1}{\partial x} &=-r\frac{R_2}{R_1}\cos(S_1-S_2)\\
		\frac{\partial S_2}{\partial t}-c \frac{\partial S_2}{\partial x} &=-r\frac{R_1}{R_2}\cos(S_1-S_2)\\
		\frac{\partial R_1}{\partial t}+c \frac{\partial R_1}{\partial x}&=rR_2\sin(S_2-S_1)\\
		\frac{\partial R_2}{\partial t}-c \frac{\partial R_2}{\partial x}&=rR_1\sin(S_1-S_2)
	\end{align}
\end{subequations}
In the regime when $S_1-S_2 \sim 0$, i.e., $\cos(S_1-S_2)\approx \sqrt{1-(S_1-S_2)^2}$, Eq. \ref{diff:dirl} and Eq. \ref{diff:dirac2} have similar solutions.  

\subsection{Non-relativistic approximation}
\label{sect:nonrel}
The Dirac equation in the non-relativistic limit reduces to the Schr\"odinger equation. Analogous approximation can be developed for Eq. \ref{AMEdIDF}. In the non-relativistic regime one expects that $R_1\sim R_2$, $S_1\sim S_2$, and $\nu=r+e$, where $e$ is the non-relativistic energy. Thus, the following expressions are employed $R_1=R+\Delta R$, $R_2=R-\Delta R$, $S_1=S-rt+\Delta S$, $S_2=S-rt-\Delta S$. Expanding Eqs. \ref{AMEdIDF} to the first order of $\Delta x$ one finds  
$\Delta S=-\frac{\Delta x}{2R}\frac{\partial R}{\partial x}$ and $\Delta R= \frac{R\Delta x}{2} \frac{\partial S}{\partial x}$.
After substituting them back and keeping all the terms up to the second order of $\Delta x$ one obtains 
\begin{subequations}
	\label{diff:schrod}
	\begin{align}
		-\frac{\partial S}{\partial t}&= -\frac{D}{2R}\frac{\partial^2 R}{\partial x^2}+\frac{D}{2} \left( \frac{\partial S}{\partial x}\right) ^2\\
		\frac{\partial R}{\partial t}&=- \frac{D}{2} \left(2\frac{\partial R}{\partial x}\frac{ \partial S}{\partial x}+ R \frac{\partial^2S}{\partial x^2}\right)
	\end{align}
\end{subequations}
which represent equations on the amplitude and phase of the wavefunction  $\psi=Re^{iS}$ of the Schr\"odinger equation ($D=\hbar/m$)
\begin{equation}
	\label{diff:schrod0}
	i\frac{\partial}{\partial t} \psi=-\frac{D}{2} \frac{\partial^2}{\partial x^2}\psi
\end{equation}
The non-relativistic approximation becomes accurate in the limit of small $\Delta x$, i.e., in the limit of fine discretization of the diffusion dynamics by a Markov chain with a large number of states - the practically interesting case. At the same time the approximation is not uniform, it breaks down in the regions where $R\rightarrow 0$ and while the wave-functions of the Schr\"odinger equation can have nodes $\psi=0$, meaning $R=0$, the exact numerical solutions of Eqs. \ref{AMEdIDF} have $R>0$, as $R=0$ corresponds to infinitely high barriers and non-ergodic dynamics in the addev sub-ensemble.

\subsection{Derivation of $q\propto e^{-W}$}
Committors for the addev sub-ensembles in the extended configuration space are defined by the following equation
\begin{subequations}
	\label{AMEq}
	\begin{align}
		u_1(i+1)[q_1(i+1)-&q_1(i)]+\nonumber  \\u_2(&i-1)[q_2(i-1)-q_1(i)]=0\\
		u_2(i-1)[q_2(i-1)-&q_2(i)]+\nonumber \\ u_1(&i+1)[q_1(i+1)-q_2(i)]=0
	\end{align}
\end{subequations}
It is straightforward to see that $q_2=q_1$, which we simply denote by $q$. The committor does not depend on the fast degree of freedom. Expressing $u_i$ as functions of $R$ and $W$, expanding the finite differences to the second order of $\Delta x$ and taking limit of $\Delta x\rightarrow 0$ one obtains the following equations on $q$: $q''+q'(W'+2R'/R)=0$, here prime denotes the spatial derivative $d/d x$. The equation is verified by $q\propto e^{-W}$, if one takes into account 
Eq. \ref{diff:schrod}b for a stationary solution $\partial R/\partial t=0$. Analogously one shows that $q'\propto e^W$. Extension to higher dimensions is discussed in appendix.

\subsection{Solving Eq. \ref{AMEdIDF} ``on the fly''.}
\label{sect:fly}
A stationary solution of Eq. \ref{AMEdIDF} can be computed ``on the fly''. For example, knowing $u_1(i)$, $v_1(i)$, $W_1(i)$, $u_2(i-1)$, $v_2(i-1)$, $W_2(i-1)$, $v_1(i-1)$, $u_2(i-2)$, and $\nu$, one can propagate the solution to the right:  from Eq. \ref{AMEdIDF}e,f one can find $u_1(i+1)$ and $v_2(i)$, from Eq. \ref{AMEdIDF}a,d one can find $W_1(i+1)$ and $W_2(i)$, from Eq. \ref{AMEdIDF}b,c one can find $u_2(i)$ and $v_1(i+1)$. Since, without the loss of generality one can set $u_1(i_0)=1$, $v_1(i_0)=1$, $W_1(i_0)=0$ for some initial $i_0$, a stationary solution of Eq. \ref{AMEdIDF} on the line has $6$ d.f..  Specifying these d.f. at one point, one is able to compute their values at any other point by propagating the solution of the equation along particle's trajectory. The values of the 6 d.f. that provide a solution that satisfies some constraints, e.g., is periodic or has zero flux, are obtained by the shooting method.

\subsection{Addev description of diffusion on the line. Interference.}
\begin{figure}[htbp]
	\centering
	\includegraphics[width=.9\linewidth]{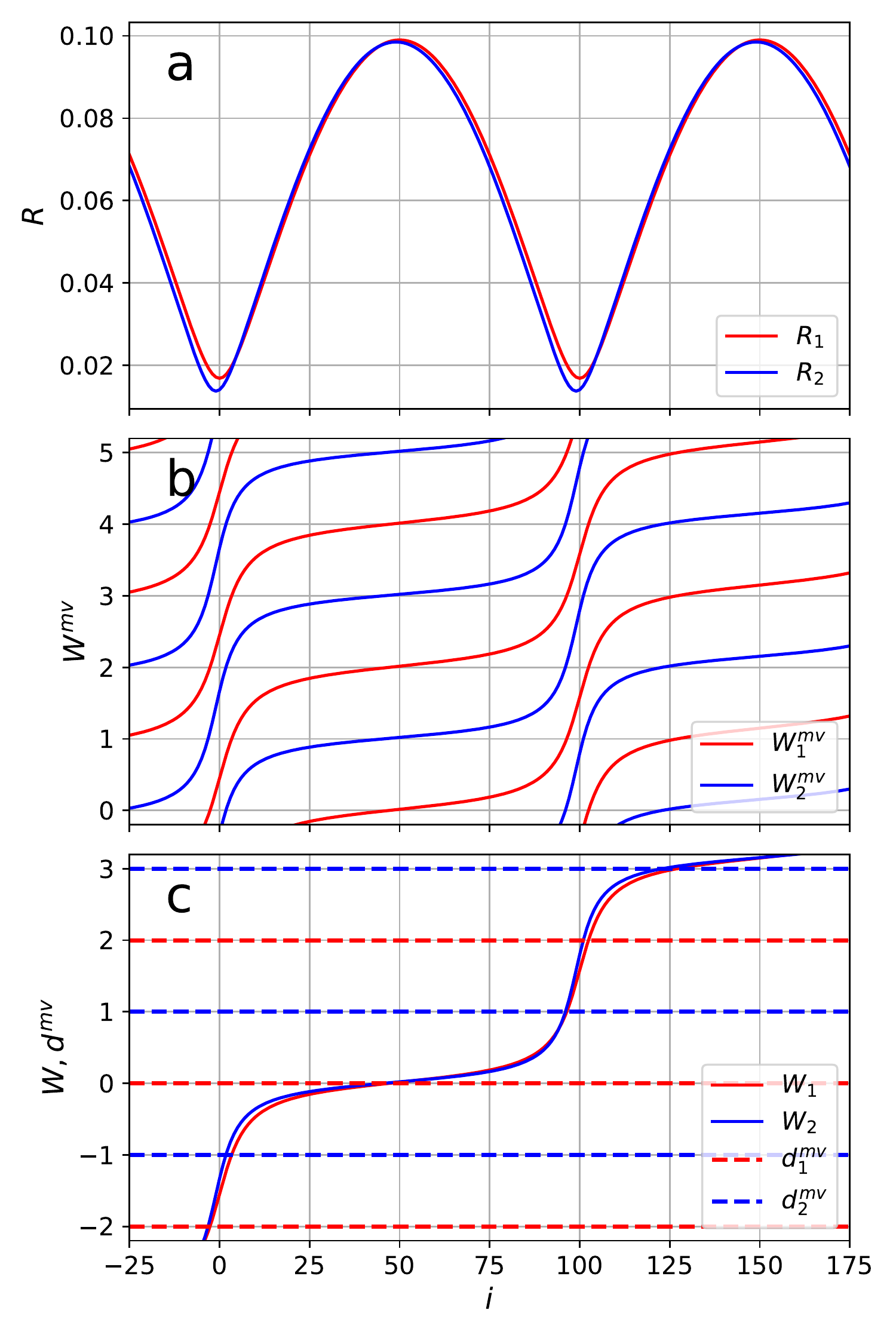}
	\caption{An addev with an internal degree of freedom for diffusion on the line. The solution is periodic. It corresponds to and can be closely approximated by a superposition of two plane waves of the Dirac equation. Panels \textbf{a} and \textbf{b} shows the amplitude $R$ and multivalued phase $W^\mathrm{mv}$ of the addev. Panel \textbf{c} shows how $W^\mathrm{mv}_j(i)=d^\mathrm{mv}_j+W_j(i)$ is decomposed into simple multivalued, independent of $x$ function $d^\mathrm{mv}_j$, and singlevalued function $W_j(i)$ of $x$. The system dynamics is described by the multivalued phase $W^\mathrm{mv}$ as follows. The system stays on the same branch, while moving in the same direction and jumps to the next branch just above at the change of the direction, $W^\mathrm{mv}_1$ and $W^\mathrm{mv}_2$ describe movement to the right and left, respectively. The simplest solution with $R_i=1$, $W_j=0$ and $\nu=r$, i.e., $W^\mathrm{mv}_j(i)=d^\mathrm{mv}_j$ describes dynamics just of the internal degree of freedom.}
	\label{fig:addev}
\end{figure}

Fig. \ref{fig:addev} shows a more complex addev solution of Eq. \ref{AMEdIDF} with periodic pattern along $x$. The solution was obtained by the shooting method, by numerically finding such 6 d.f. specifying the solution, which are reproduced after 100 steps, e.g., $u_j(i)=u_j(i+100)$, here and below we assume $r=1$, $\Delta x=1$; the 6 d.f. were initialized by the superposition of two plane-waves of the Dirac equation \ref{diff:dirac2}.  

We will analyse this solution in detail to illustrate how the addev solutions of Eq. \ref{AMEdIDF} describe stochastic dynamics. The solution can be closely approximated by the corresponding solution of the Dirac equation, describing a superposition of two plane waves; the difference between the solutions of Eqs. \ref{AMEdIDF}, \ref{diff:dirl} and \ref{diff:dirac2} is less than 1\%. Note that, strictly speaking, Eq. \ref{AMEdIDF} is non-linear and a superposition of two solutions is ill-defined. However, when it can be approximated by the linear equations of QM, one can approximately talk about superpositions. Thus, Fig. \ref{fig:addev} shows a superposition of two addevs corresponding to plane waves. The superposition shows periodic oscillations in the amplitude - an evidence of two waves interfering.

Figs. \ref{fig:addev}b and \ref{fig:addev}c show how the complex function $W^{mw}=d^\mathrm{mv}+W$ is decomposed on two simple functions $d^\mathrm{mv}$ and $W$. Namely, the complex pattern of the multivalued phase function $W^\mathrm{mv}$ (Fig. \ref{fig:addev}b), is obtained simply by translating $W_1$ and $W_2$ vertically with constant step of $2$, which is due to the contribution of the $d^\mathrm{mv}$ component. In the region of large values of $R$, $W_i$ changes little and are roughly linear functions of $x$, and alternatively, $W_i$ change significantly and in a highly non-linear manner as $R$ get smaller. The stochastic dynamics is projected on $W^\mathrm{mv}$ in the following way: while the system moves to the right $i\rightarrow i+1$, it stays on a branch corresponding to $W^\mathrm{mv}_1$, with change $\Delta W^\mathrm{mv}=W^\mathrm{mv}_1(i+1)-W^\mathrm{mv}_1(i)=W_1(i+1)-W_1(i)$, as soon as it makes a step in the opposite direction $i\rightarrow i-1$, it jumps to the next the branch $W_2^\mathrm{mv}$ just above, with change $\Delta W^\mathrm{mv}=W^\mathrm{mv}_2(i-1)-W^\mathrm{mv}_1(i)=1+W_2(i-1)-W_1(i)$ and so on. Thus, at every change of direction the system jumps one branch up, which leads to steady rapid growth of $W^\mathrm{mv}$.

One can describe the addev dynamics at two time-scales, fast and slow. At fast time-scale, the dynamics is described using $W^\mathrm{mv}$ (Fig. \Ref{fig:addev}b). Since ($R$, $W^\mathrm{mv}$) satisfy the AME, the dynamics of an addev sub-ensemble projected on phase $W^\mathrm{mv}$ is simple - trajectories move with constant drift speed of $\nu$. Flux along $W^\mathrm{mv}$ is constant and satisfies $J_{W^\mathrm{mv}}=\nu/\Delta_\odot W^\mathrm{mv}=\nu/2$. However, dynamics along $W^\mathrm{mv}$ can not be described as diffusion, it contains large jumps of length $\Delta W^\mathrm{mv}\approx 1$, when the system changes direction. The coordinate $W^\mathrm{mv}$, which is the sum of $d^\mathrm{mv}$ and $W$ mixes the description of the very fast dynamics of the internal degree of freedom ($d^\mathrm{mv}$) with the slow configuration dynamics ($W$), which is of main interest. 

At the other, slow time-scale, the dynamics is described by $W$. $W$ describes just the slow configuration dynamics and separates it from the mixture. $W$, by itself, however, is not the phase of an addev, and thus trajectories along it do not move with constant drift speed. When $W_1\approx W_2$ (Fig. \Ref{fig:addev}c) the changes of $W$ during each step become very small and dynamics can be approximated by a diffusive model similar to that on Fig. \ref{fig:FW}. In the limiting case of small $\Delta x$ when $W_1\approx W_2$, subtracting $rt$ from $\nu t$,  the leading contribution that comes from the internal dynamics, one obtains Eq. \ref{diff:schrod}, which is the Schr\"odinger equation. 

Since the rate matrix in the extended configuration space does not satisfy the detailed balance, the committors are not defined by $q\propto e^{-W^\mathrm{mv}}$. Since $W^\mathrm{mv}$ mixes fast internal dynamics with slows spatial dynamics, we are more interested in committors as functions of configuration space, where we have $q\propto e^{-W}$. In fact, the fast internal dynamics is automatically factored out since $q_1(i)=q_2(i)$. This suggests a generic property that committors average the fast internal dynamics out and are functions of slows dynamics only, described by $W$. 

Such a separation of the fast and slow dynamics can be also interpreted as an example of coarse-graining of addev dynamics, which, as described in section \ref{sect:potential} may be a source of potential. Indeed, here the coarse-graining essentially lumps microstates with different value of the internal degree of freedom together. The micro dynamics within a macrostate contributes to the grows of $S$ with rate $\approx rt$, which corresponds to the effective potential of $U=r$. Removing this large constant potential from the description one obtains Eq. \ref{diff:schrod}. 

The stochastic dynamics in the addev sub-ensemble is completely specified by the biased rate matrix $\tilde{K}(i|j)$, which is known once an addev solution is found. Having the matrix at hand, one can compute any properties of the stochastic dynamics. However, many useful properties of the stochastic dynamics can be computed just from the module and phase ($R$, $W$) of the addev, or from the approximating wave-function. For example, the exact expression for the flux along $x$, Eq. \ref{J}, takes the form
\begin{equation}
	\label{Jd}
	\begin{aligned}
	J=r[u_1(i+1)v_1(i)+u_1(i+1)v_2(i)\\ -u_2(i)v_1(i+1)-u_2(i)v_2(i+1)]
	\end{aligned}
\end{equation}
The expression is a bilinear function, and in the non-relativistic limit can be approximated by the familiar expression of $DR^2\partial W/\partial x=D\mathfrak{Im}(\psi^* \partial \psi/\partial x)$, recall from section \ref{sect:nonrel} that $D$ corresponds to $\hbar /m$. For the superposition of two plane waves moving in the opposite directions, with amplitudes $a$ and $b$, considered here, the flux (along $x$) is equal $J_0(|a|^2-|b|^2)$, where $J_0$ is the flux for a single wave; i.e., (the inclusion of) the wave moving in the opposite direction decreases the flux. The addev/diffusive model interpretation of the smaller value of the flux is following: the flux is smaller because the diffusive dynamics now proceed on the free energy profile with the barriers ($R$ is non-constant), and the higher are the barriers, the smaller is the flux. Note that flux along $x$ is different from that along $W^\mathrm{mv}$ which is the same for all the superpositions with the same $\nu$. Since the phase $W$ defines the committor functions, e.g., $q\propto e^{-W}$, one can compute such properties of the dynamics as the mfpt, the mtpt, etc.

\subsection{Addev description of diffusion on the interval. Bound states.}
\label{sect:bound}
Next we consider addev solutions of Eq. \ref{AMEdIDF} on the interval, e.g., for a finite Markov chain,  corresponding to QM bound states. Such a solution has zero flux and satisfy $-W_2(i-1)=W_1(i)$, $u_2(i-1)=v_1(i)$ and $v_2(i-1)=u_1(i)$. 
A solution for a finite Markov chain can be specified by requiring probabilities outside the Markov chain being zero: $v_1(-1)=0$ and $u_1(N-1)=0$. It reduces the number of d.f. to 2: $W_1(0)$ and $\nu$, from which the other d.f. are found as $u_1(0)=v_2(-1)=1$, $v_1(0)=u_2(-1)=(1+2W_1(0))/\nu$, $v_1(-1)=u_2(-2)=0$, $W_2(-1)=-W_1(0)$. The solution is obtained by the shooting method by finding such $W_1(0)$ and $\nu$ so that $u_1(N-1)=0$. 

\begin{figure}[htbp]
	\centering
	\includegraphics[width=.9\linewidth]{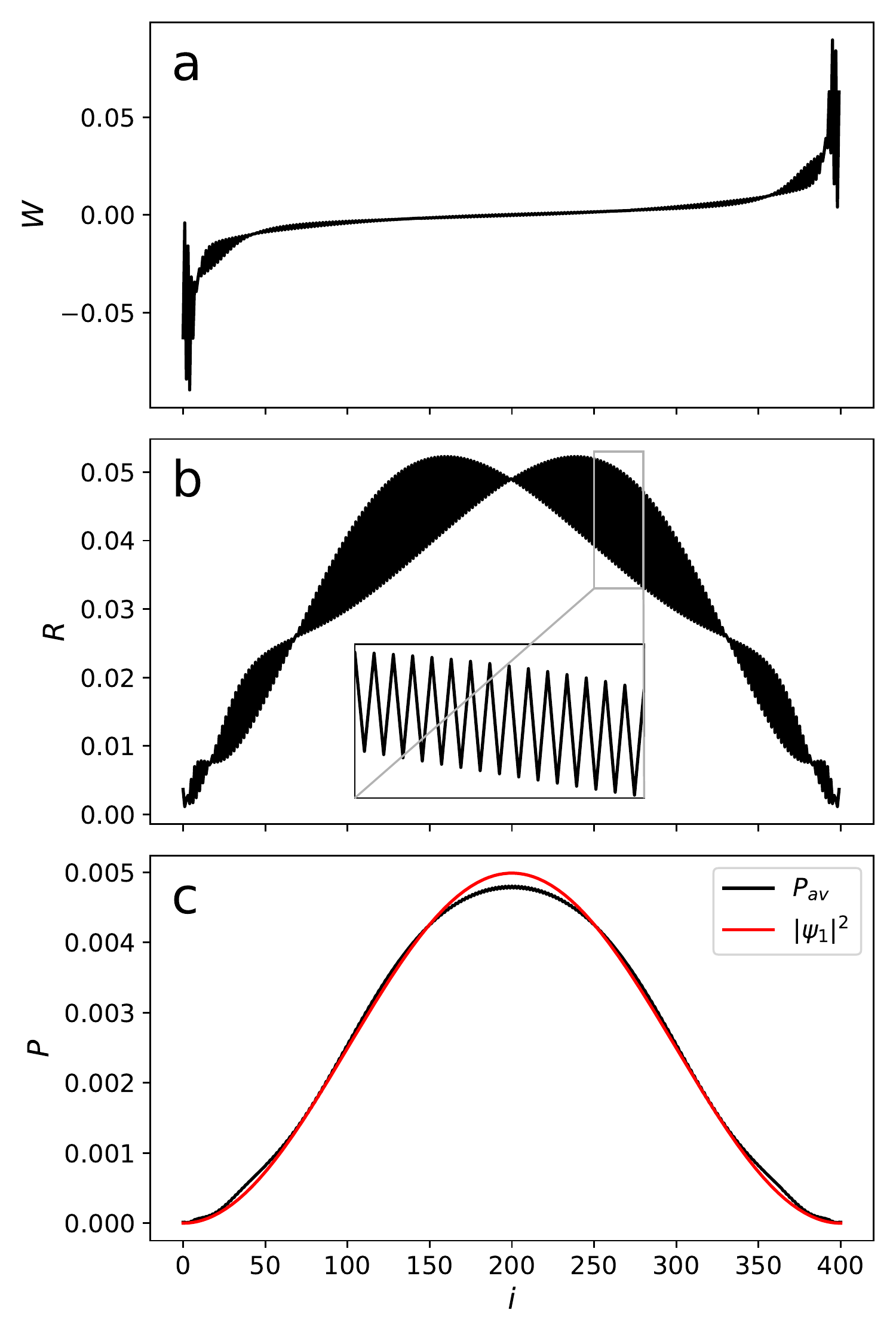}
	\caption{An addev with an internal degree of freedom for diffusion on the interval. The solution satisfy the following relations: $-W_2(i-1)=W_1(i)$, $R_2(i-1)=R_1(i)$, which are denoted as $W(i)$, $R(i)$ and shown on panels \textbf{a} and \textbf{b}, respectively. Inset in panel (b) shows the sawtooth shape of the solution. Panel (c) shows that the  stationary probability, averaged over the internal degree of freedom, $P_{av}(i)=R_1(i)^2+R_2(i)^2$ (black line) is in very good agreement with that of the corresponding solution of the Schr\"odinger equation  $|\psi_1|^2(i)\sim \sin^2(\pi i/N)$ (red line). This solution has zero flux and corresponds to the quasi-stationary distribution shown on Fig. \ref{fig:TPtau}}
	\label{fig:well}
\end{figure}

Fig. \ref{fig:well} shows an addev solution for a Markov chain with $N=401$ states.
The solution demonstrates a qualitatively new property: it is not smooth, but have a sawtooth profile (inset in panel \textbf{b}). Such solutions can not be described by the continuous Eq. \ref{diff:dirl}. The probability as a function of the spatial coordinate, which is obtained by averaging over the internal degree of freedom, $P_{av}(i)=R_1^2(i)+R_2^2(i)$, is in a good agreement with that of the corresponding solution of the Schr\"odinger equation $|\psi_1|^2(i)=2/N\sin^2(k_1x)=2/N\sin^2(\pi x/L)$ for $k_1=\pi/L$, $x=(i-1)\Delta x$ and $\Delta x=L/(N-1)$, where $L$ is the well width. This probability does not have the sawtooth shape as the maxima and minima overlap exactly since $R_2(i-1)=R_1(i)$. 

The flux for such a bound state solution is exactly zero. Hence, it can not be used for the description of the dynamics as a stochastic periodic process of the system moving between the two boundaries analogous to that described in sections \ref{sect:diff_segm} and \ref{sect:potential}. Also, the phase of the addev $W$ is close to zero, and in the non-relativistic limit it approaches zero similar to the bound states wave-functions of the Schr\"odinger equation, which are real.

This solution was described because of the following reasons. First, it is an example of an addev approximating the quasi-stationary distribution, i.e., the sub-ensemble of trajectories conditioned on never reaching the two boundary states, which approximates the TPs with very long duration shown on Fig. \ref{fig:TPtau}. Second, it demonstrates that the AME allows bound state solutions, e.g., that in an infinite square well. Note that existence of bound state solutions for both Eqs. \ref{diff:dirl} and \ref{diff:dirac} is a non-trivial question. The PDE Eq. \ref{diff:dirl} is not suitable for the description of bound states, i.e., states where $R_j(x)\rightarrow 0$ when $x\rightarrow \pm \infty$. From Eqs. \ref{diff:dirl}cd one obtains $\partial/\partial x (R_1^2-R_2^2)=0$, meaning that $R_1(x)=R_2(x)$ and $W_1(x)=-W_2(x)$. It means that difference $W_1(x)-W_2(x)=2W_1(x)$ can grow larger than $1$, e.g., for excited states, while formalism of Eqs. \ref{AMEdIDF} and \ref{diff:dirl} requires that $|S_1-S_2|=|W_1-W_2|<1$. Also, specification of the boundary conditions (BC) is not straightforward. Consider, for example, stationary solutions for the Dirac equation in an infinite square well \cite{alonso_general_1997}. BC taken as $\psi_1(0)=\psi_2(0)=\psi_1(L)=\psi_2(L)=0$, where $L$ is the width of the well, allow only the trivial solution. More general BC can be developed based on self-adjoint extensions of symmetric
operators \cite{alonso_general_1997}, however it is not clear how this can be applied to non-linear Eq. \ref{diff:dirl}.

\subsection{Bound states addevs with non-zero flux} 
To describe dynamics via an addev as a periodic process we need to construct an addev solution with a non-zero flux which goes from one boundary to the other and back. One may proceed analogously to section \ref{sect:diff_segm}, by taking the plane wave solution and reflecting it at the boundaries. For the forward path $u_1(x,t)=v_1(x,t)=\sqrt{\nu+ck}$, $u_2(x,t)=v_2(x,t)=\sqrt{\nu-ck}$, $S_1(x,t)=S_2(x-\Delta x,t)=W(x)-\nu t$ and $W(x)=2jkL+kx$; for the backward path $k$ is replaced by $-k$ and, correspondingly,  $u_1=v_1$ are exchanged with $u_2=v_2$ and $W(x)=2(j+1)kL-kx$, here we use continuous variable $x=i\Delta x$ and $j$ indicates different branches for the forward and backward paths analogous to that in section \ref{sect:diff_segm}, i.e., $W$ is a multivalued function. Such an addev describes a sub-ensemble of trajectories performing biased random walk towards one boundary, then in the opposite direction and so on. This solution can also be obtained by  computing it ``on the fly''. Starting with plane wave, one propagates the solution till it reaches the boundary, where the solution is reflected and so on. Alternatively, as described in section \ref{sect:diff_segm}, the forward and backward pathways can be joined into a circle, and one considers a plane wave addev solution on the circle with $W(x)=kx$, where $x$ and, correspondingly, $W$, are multivalued coordinates that cover the circle repeatedly, analogous to the angle. A constant bias is applied along the entire circle, with the diffusive model corresponding to that on Fig. \ref{fig:FW}. 

Consider the ``on the fly'' approach to constructing the solution. In principle, instead of a single plane wave one may also propagate a superposition of two interfering plane waves with equal and opposite momenta, resulting in an interfering plane wave running between the boundaries. The solution has a non-zero flux, and, in principle, it can be used for the description of reaction dynamics. We note however, that in approximating reaction dynamics with addevs it is reasonable to seek such addevs, which captures most of the reaction flux, meaning that a reaction can be described with a small number of such addevs. Thus, such solutions with interfering plane waves which have smaller flux are suboptimal and not considered for the description of reaction dynamics. 

Interestingly, the probability to observe an interfering plane wave, computed using KL divergence, equals $e^{-A\Delta t}$, where $A=r(r/\nu-1)[|a|^2+|b|^2+\min(|a|^2,|b|^2)]$, where $a$ and $b$ are the waves amplitudes. It means that two non-interfering waves are more likely to be observed then two interfering.

\begin{figure}[htbp]
	\centering
	\includegraphics[width=.9\linewidth]{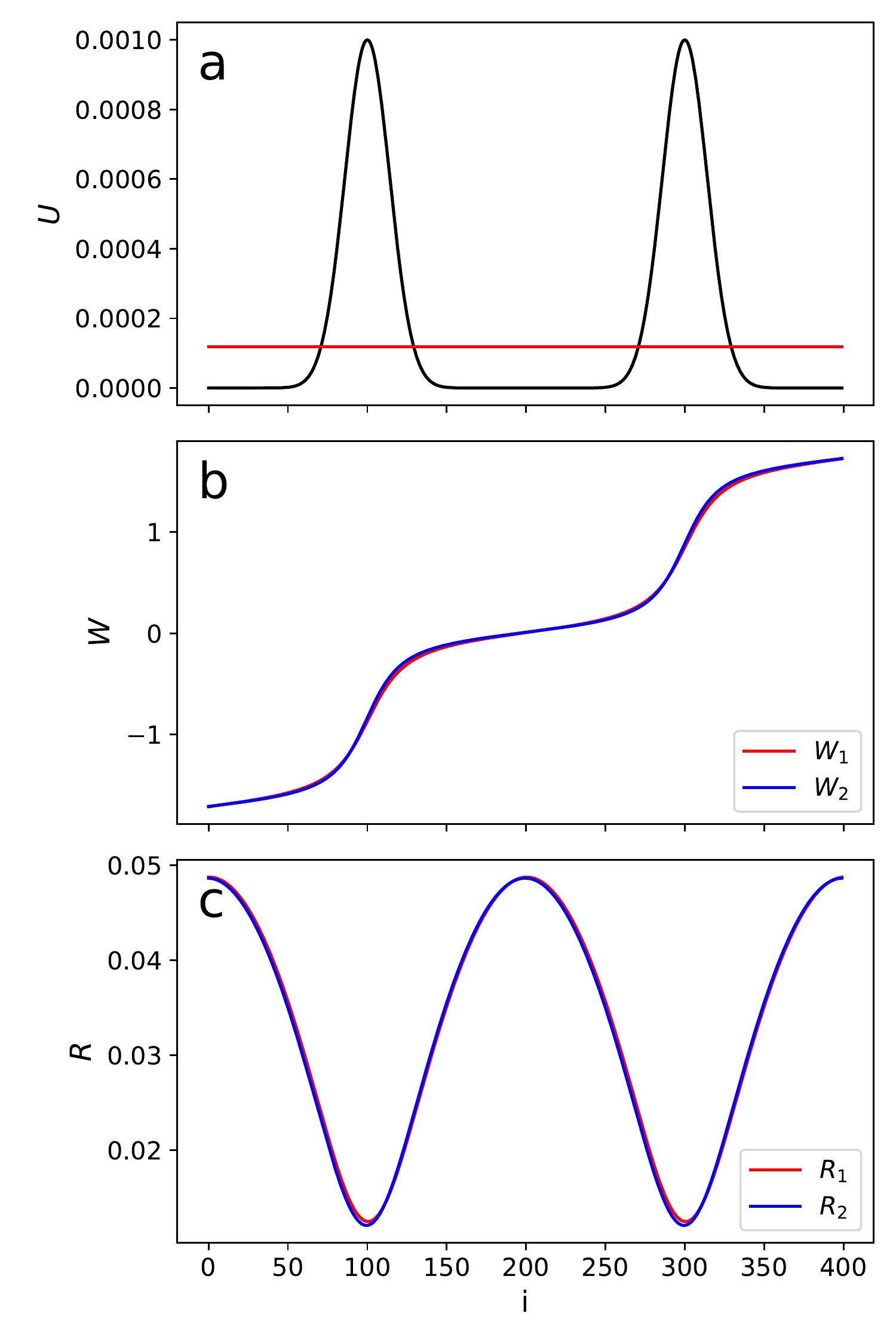}
	\caption{An addev with an internal degree of freedom for diffusion on the interval over a barrier with a non-zero flux. The system is modelled as diffusion on the circle with explicit representation of forward (first half) and backward (second half) paths. Potential $U$, phase $W_i$, and amplitude $R_i$ as functions of the coordinate along the circle are shown on panels \textbf{a}, \textbf{b}, and \textbf{c}, respectively. The energy $\nu-r$ (red line on panel \textbf{a}) is lower than the barriers. Phase $W_i$ is a multivalued function, in contrast to Fig.\ref{fig:well}a.}
	\label{fig:wellU}
\end{figure}

\subsection{Bound states addevs with non-zero flux and potential}
Finally, we construct a bound state with non-zero flux addev solution for diffusion over a barrier. The potential in Eq. \ref{AMEdIDF} is introduced  according to Eq. \ref{AMEU} by replacing $-dS_j(i)/dt$ with $-dS_j(i)/dt-U(i)$. This leads to the expected results of scalar potential $\nu-U$ in Eq. \ref{diff:dirl} and $e-U$ in Eq. \ref{diff:schrod} or the standard term $U\psi$ in Eq. \ref{diff:schrod0}. The potential can be also introduced as a variable rate $r=r_0+U$, which leads to the same non-relativistic limit.

Fig. \ref{fig:wellU} shows an addev solution for the diffusion on the interval over a barrier, as a model of reaction dynamics. We use the same construction of the addev on the circle. The circle contains two barriers (Fig. \ref{fig:wellU}a), one for the forward pathway (first half) and one for the pathway back (second half).  The probabilities are exponentially small at the top of the barriers (Fig. \ref{fig:wellU}c) and are large at the walls, i.e., they high where potential is low and vice-versa in agreement with the canonical distribution and in contrast with the addev solutions corresponding to classical mechanics (section \ref{sect:classical}). The phase function $W$ while being locally singlevalued, becomes multivalued globally, it increments after the full trip from one boundary to the other and back. The energy of the solution $\nu-r$ is smaller than the barrier height (Fig. \ref{fig:wellU}a), meaning that such a solution describes the desired fluctuation of energy and transition over the barrier. This solution corresponds to the QM tunnelling. The solution has a non-zero flux and describes the stochastic reaction dynamics as a periodic stochastic motion between the walls - a stochastic eigenmode.

\section{Concluding Discussion}
The paper continues the development of the framework of additive eigenvectors, generalizing the optimal RCs to non-equilibrium dynamics. Here we consider stationary solutions of the addev master equation and illustrate the developments by analysing diffusion. We have shown that forward and time-reversed committors can be computed as $q\propto e^{-\alpha W}$ and $q'\propto e^{\alpha W}$ for systems with the detailed balance, and for practically interesting non-relativistic limit of diffusion. It means that a diffusive model along an optimal RC ($W$) can be used to compute exactly many important properties of the sub-ensemble dynamics. We have described a self-consistent way to introduce interaction potential into the addev framework.  We have demonstrated that stochastic dynamics in an addev can be efficiently sampled by solving the addev master equation ``on the fly''. It means that an addev stochastic dynamics can be defined and sampled by itself, not as a sub-ensemble of trajectories of a Markov chain. We have described two families of addev solutions for diffusion, that without and with an internal degree of freedom. We have shown that the latter provide the desired description of stochastic reaction dynamics as a stochastic periodic process - a stochastic eigenmode.

One common assumption in the analysis of molecular dynamics simulations is that the dynamics is Markovian in the configuration space. It is shared by many approaches, including the free energy landscapes and the Markov state models. However, the dynamics is Markovian in the phase space and becomes approximately Markovian in the configuration space if observed at relatively large time-scales, when the system forgets its momenta. This, in particular, introduces a lower bound on the lag times or lengths of trajectories that can be analysed by such methods. Which, correspondingly, limits efficiency of parallel approaches for exascale computing, where one uses a very large ensemble of short trajectories instead of a single long one \cite{lohr_abeta_2021, kohlhoff_cloud_2014, zhou_fegs_2012, doerr_learning_2014, wan_adaptive_2020, perez_adpative_2020, pan_string_2008, lev_string_2017}. Accurate analysis of classical dynamics in phase space is one motivation for the development of the addev framework. One approach to analyse such dynamics is provided by the Mori-Zwanzig formalism, where dynamics, projected on an arbitrary RC, is described by a generalized Langevin equation with a memory kernel \cite{zwanzig_memory_1961, mori_transport_1965}. While the approach may provide a quantitatively accurate description \cite{lei_data_2016}, it makes the interpretation and  understanding of the dynamics difficult. What is the meaning of the free energy landscape or the transitions state (the bottleneck of the reaction) when the memory effects are important? Another approach models the molecular dynamics with the standard, memory-less Langevin equation, and computes for the latter the forward and time-reversed committors, which allows one to compute exactly many important properties of the dynamics between the boundary states \cite{e_towards_2006,lu_exact_2014}. However, since the two committors are generally different functions it is not clear how to combine them into a single optimal RC. The formalism developed here provides a different solution. The classical dynamics in the phase space can be closely approximated by an addev dynamics, which simultaneously provides a single RC optimal for forward and time-reversed dynamics. The diffusive model along the optimal RC (the action function) can be used to compute the forward and time-reversed committors and thus to compute exactly many important properties of the dynamics. The diffusive model describes the dynamics as non-equilibrium diffusion with non-zero flux on a multivalued free energy profile (see Fig. \ref{fig:FW}).

Analogous conclusions can be made regarding addevs approximating QM systems. In the regime, where the approximation is close, the stochastic dynamics of the addev sub-ensemble provides a stochastic model of the corresponding stationary wave-function. The addev dynamics, in turn, is described by the diffusive model along the optimal RC (the addev or wave-function phase). The ability to solve the AME ``on the fly'' suggests the possibility of efficiently sampling such addevs and the corresponding QM wave-functions, where the approximation is close. 
The addev equations are different from the QM equations, e.g., the former are non-linear, meaning that the addevs are not an interpretation of QM. However, there are obvious similarities between the two, which can be used to guide the development of the former and provide stochastic models for the latter.

The examples described here are rather elementary, however they are sufficient to illustrate the framework. An addev defines a sub-ensemble of trajectories and a corresponding optimal RC. The stochastic dynamics in the sub-ensemble is described by the Markov chain with the biased rate matrix $\tilde{K}$, which can be used to compute any property of the dynamics. At the same time, one can approximate the dynamics by a diffusive model $F(W)$ and $D(W)$, as shown on Fig. \ref{fig:FW}. The dynamics is non-equilibrium with a non-zero flux. It describes a stochastic periodic process, a stochastic eigenmode. If $W$ is the phase of an addev, i.e., $Z_{C,1}(W)$ is constant, $F(W)$ can be determined from the stationary probability $P(W)$ using Eq. \ref{Fw}. If not, $F(W)$ can be determined by integrating $\int dW/Z_{C,1}(W)$ \cite{krivov_addev_2022}. However, since the forward and time-reversed committors are both functions of $W$, one can compute many important properties of the dynamics exactly just from the addev's $R$ and $W$ or from the corresponding  wave-function, without constructing the actual model of dynamics. Some of the corresponding functionals are bilinear and take a familiar expression when acting on a wave-function. The higher dimensional systems are obviously more complex, however the complexity is mainly due to the optimal RC $W$ being a complex multidimensional function, projecting a multidimensional configuration space onto a circle. The diffusive model, i.e., $F(W)$ and $D(W)$, is still one-dimensional and describes dynamics in the same way. The situation is analogous to the description of the dynamics with the committor RC, where the most difficult part is the accurate determination of the complex projection performed by the committor function \cite{krivov_protein_2018}.

Usually, stochastic processes/trajectories are conditioned on a specific function of the trajectory having some value \cite{Evans_rules_2004, hedges_dynamic_2009, Garrahan_first_2009, jack_large_2010, Chetrite_Nonequilbrium_2015, Derrida_large_2019}, e.g., $A_T=a$, where  $$A_T=\frac{1}{T}\int_0^T f(X_t) dt+\frac{1}{T}\sum_{0\le t\le T: X_{t^-}\neq X_{t^+}} g(X_{t^-},X_{t^+}),$$ here $X_t$ is trajectory, functions $f$ and $g$ describe contributions of the states and transitions of a Markov chain, respectively \cite{Derrida_large_2019,Chetrite_Nonequilbrium_2015}. Such observables include many variables of mathematical (occupation time, number of transitions) and physical (entropy production, currents, activity, work) interest, see Ref. \citenum{Chetrite_Nonequilbrium_2015} and references therein. The addev conditioning does not focus on any such specific quantity. It requires the addev sub-ensembles to have the same optimal RC for forward and time-reversed dynamics and while being a reasonable requirement for a RC, may seem rather ad hoc and abstract for stochastic processes and for physics in general. However, the two families of addev solutions obtained here correspond to the fundamental equations of classical and quantum mechanics, suggesting that addevs define a class of conditioned stochastic processes suitable for the description of dynamics of physical origin.

The work presented here can be continued in the following directions.
Extension of the non-parametric RC optimization approach \cite{banushkina_nonparametric_2015, krivov_protein_2018, krivov_blind_2020, krivov_nonparametric_2021} to the described addev solutions for diffusion. In particular, it should allow the analysis of large ensembles of short molecular dynamics trajectories. Analysis of realistic, relatively complex model systems. Development of addev solutions with other constructions of internal degrees of freedom. Development of efficient sampling methods, especially for addevs with internal degrees of freedom in higher dimensions.

\appendix*
\section{Generalisations to many dimensions}
Consider a Markov chain describing isotropic diffusion in d-dimensional space with rate matrix $K(\vec i \pm \vec{e}_j|\vec{i})=r$, where $r=D/\Delta x^2$, $\vec{i}$ denote points on the d-dimensional hypercube lattice with spacing $\Delta x$ and $\vec{e}_j$ is the unit vector along $j$-th coordinate. For simplicity we consider isotropic diffusion with constant diffusion coefficient. The AME (Eq. \ref{AMEU}) in the limit of $\Delta x \rightarrow 0$ can be written as the following PDE
\begin{subequations}
	\label{addevd_dim}
	\begin{align}
		D\sum_j \frac{\partial}{\partial x_j}\left (P\frac{\partial S}{\partial x_j}\right)=& 0\\
		-D\sum_j \frac{\partial S}{\partial x_j} \frac{\partial \ln q}{ \partial x_j}+U = & -\frac{\partial S}{\partial t}\\
		D\sum_j \frac{\partial}{\partial x_j}\left(P\frac{\partial \ln q}{\partial x_j}\right)=& \frac{\partial P}{\partial t},
	\end{align}
\end{subequations}
where $q=v/u$.

Consider now a stationary solution $\frac{\partial P}{\partial t}=0$ and $S=W-\nu t$. Eqs. \ref{addevd_dim}a and \ref{addevd_dim}c have similar form, but it does not follow that $W$ and $\ln q$ are simply related, because they may have different boundary conditions. Note, however, than in considering dynamics in the addev sub-ensembles, one is not interested in any committor function, but in the committor function of trajectories projected on the optimal RC $W$, it means that boundary conditions for $W$ and $q$ are simply related, i.e., iso-surfaces $q=q_A$ and $W=W_A$ coincide, and the same for $q=q_B$ and $W=W_B$ (see section \ref{sect:q}). For these solutions, $\ln q \propto -\alpha W$, and normalizing $W$ to $\alpha =1$, one obtains the Hamilton-Jacobi equation for the action function. 
\begin{equation}
	\label{addevd5}
		D\sum_j \left(\frac{\partial S}{\partial x_j}\right)^2+U = -\frac{\partial S}{\partial t}\\
\end{equation}
Multidimensional analogues of Eqs. \ref{pdeaddev1d} and  \ref{lanj} are
\begin{subequations}
	\label{pdeaddevmd}
	\begin{align}
		\frac{\partial P}{\partial t}=&-\sum_i \frac{\partial J_i}{\partial x_i} \\
		J_i=&-D\frac{\partial P}{\partial x_i} + DP\frac{\partial \ln(u^2)}{\partial x_i}
	\end{align}
\end{subequations}
and
\begin{subequations}
	\label{lanjmd}
	\begin{align}
	\Delta x_i=&D\Delta  t \left(p_i-\frac{f_i}{2(\nu-U)}\right)+\xi_i\sqrt{2D \Delta t}\\
	\Delta p_i=&\frac{f_i}{2D}\frac{\sum_i f_i\Delta x_i}{\sum_i f_i p_i},
	\end{align}
\end{subequations}
where $\Delta x_i=x_i(t+\Delta t)-x_i(t)$, $\Delta p_i=p_i(t+\Delta t)- p_i(t)$ and  $x_i$, $p_i=\frac{\partial W}{\partial x_i}$, $f_i=-\frac{\partial U}{\partial x_i}$, and $\xi_i$ are vectors of the position, momenta, force and independent normally distributed random variables, respectively. Eq. \ref{lanjmd}b updates momenta or the gradient of $W$ ``on the fly'' and is obtained by assuming that only the momenta component parallel to the force changes, and by amount determined from the conservation of energy. A subtle point is integrating the solution around the turning points, i.e., the jumps into the classically forbidden regions; one approach is to reflect the components of $\Delta x$ and $\Delta p$ parallel to the force.

Next we want to show that for diffusion in $d$ dimensions, addevs with an internal degree of freedom  are approximated, in the practically important non-relativistic limit, by the $d$-dimensional Schr\"odinger equation and that $q\propto e^{-W}$. However, generalisation of such addevs to many dimensions is much more complicated. Finding a satisfactory generalisation of the internal degrees of freedom to higher dimensions, which in particular respects the spherical symmetry, is a work in progress. Thus the results below should be considered as a strong conjecture. We assume that non-relativistic limits, which completely decouple fast internal dynamics and slow configuration dynamics of interest, should all look very similar, thus one can take any simple, probably not very physically realistic system and study its non-relativistic limit. Obviously, for the description of the limits or corrections containing traces of the fast dynamics one needs to know  the fast dynamics.

First, we note a useful addev property, the separability. Namely, it is straightforward to verify that for a system, consisting of two independent subsystems, solutions of the AME (Eq. \ref{AMEU}) factorize $u(i_1,i_2,t)=u_1(i_1,t)u_2(i_2,t)$, $v(i_1,i_2,t)=v_1(i_1,t)v_2(i_2,t)$, $R(i_1,i_2,t)=R_1(i_1,t)R_2(i_2,t)$, $S(i_1,i_2,t)=S_1(i_1,t)+S_2(i_2,t)$, $U(i_1,i_2)=U_1(i_1)+U_2(i_2)$ and for a stationary state $\nu=\nu_1+\nu_2$, or for the approximating wave-functions $\psi(i_1,i_2)=\psi_1(i_1)\psi_2(i_2)$; here subscripts $1$ and $2$ refer to the quantities of the two subsystems. Next, we model $d$-dimensional diffusion or $d$-dimensional random walk, described by the Markov chain as $d$ independent one-dimensional random walks. The addev for d-dimensional diffusion factorizes on addevs describing diffusion along the corresponding coordinate ($j$-th), which can be approximated in the non-relativistic regime by the corresponding Schr\"odinger equation
\begin{equation}
	\label{diff:schrod0_dim}
	i\frac{\partial}{\partial t} \psi_j=-\frac{D}{2} \frac{\partial^2}{\partial x_j^2}\psi_j+U_j(x_j)\psi_j
\end{equation}
where $\psi_j$ is a function of $t$ and $x_j$. Combining independent wave-functions together as $\psi=\prod_j \psi_j$ one obtains the d-dimensional Schr\"odinger equation, which approximates the addev for d-dimensional diffusion.  

\begin{equation}
	\label{diff:schrod1_dim}
	i\frac{\partial}{\partial t} \psi=-\frac{D}{2} \left(\sum_j \frac{\partial^2}{\partial x_j^2}\right)\psi +U\psi
\end{equation}

Analogously, one can show that the committor function for a system, consisting of two independent subsystems, factorizes as $q(i_1,i_2)=q_1(i_1)q_2(i_2)$, and obtain the following equation for the high-dimensional committor

\begin{equation}
\sum_j \frac{\partial^2 q}{\partial x_j^2} +\frac{\partial q}{\partial x_j} \left(\frac{\partial W}{\partial x_j} +2\frac{\partial \ln R}{\partial x_j}\right)=0
\end{equation}
The equation is verified by $q\propto e^{-W}$, if one takes into account $$\sum_j 2\frac{dR}{dx_j}\frac{ dW}{dx_j}+ R \frac{d^2W}{dx_j^2}=0,$$ the multidimensional form of  Eq. \ref{diff:schrod}b, obtained from Eq. \ref{diff:schrod1_dim}. 
This agrees with simple argument that since for each independent random walk $q_j\propto e^{-W_j}$, for their product one has $q\propto e^{-W}$.

Note that isotropic d-dimensional diffusion is spherically symmetric and should be invariant to arbitrary rotations of the coordinate system, which means that the final equations should be rotationally invariant; the derived equations are. Combination of the rotational symmetry with the requirement to describe the independent random walkers impose strong constraints on the equations, suggesting that the derived equations are correct generalisations to higher dimensions. 


%
\end{document}